\documentclass[iop]{emulateapj}
\usepackage{epsfig}
\usepackage{amsmath}
\usepackage{natbib}

\providecommand{\e}[1]{\ensuremath{\times 10^{#1}}}

\slugcomment{To appear in Astrophysical Journal}

\shorttitle{Flux Tubes in Fully Convective Stars}
\shortauthors{Weber \& Browning}

\begin{document}


\title{Modeling the Rise of Fibril Magnetic Fields in Fully Convective Stars}


\author{Maria A. Weber and Matthew K. Browning}
\affil{Department of Physics and Astronomy, University of Exeter, Stocker Road, EX4 4QL Exeter, UK}



\begin{abstract}
Many fully convective stars exhibit a wide variety of surface magnetism, including starspots and chromospheric activity.  The manner by
which bundles of magnetic field traverse portions of the convection zone to
emerge at the stellar surface is not especially well understood. In the Solar
context, some insight into this process has been gleaned by regarding the
magnetism as consisting partly of idealized thin flux tubes (TFT).  Here,
we present the results of a large set of TFT simulations in a rotating
spherical domain of convective flows representative of a 0.3M$_{\odot}$
main-sequence star.  This is the first study to investigate how individual flux tubes in such a
star might rise under the combined influence of buoyancy, convection, and
differential rotation.  A time-dependent hydrodynamic convective flow
field, taken from separate 3D simulations calculated with the anelastic
equations, impacts the flux tube as it rises. Convective motions modulate
the shape of the initially buoyant flux ring, promoting localized rising
loops.  Flux tubes in fully convective stars have a tendency to rise nearly
parallel to the rotation axis. However, the presence of strong differential
rotation allows some initially low latitude flux tubes of moderate strength
to develop rising loops that emerge in the near-equatorial region.
Magnetic pumping suppresses the global rise of the flux tube most
efficiently in the deeper interior and at lower latitudes.  The results of
these simulations aim to provide a link between dynamo-generated magnetic
fields, fluid motions, and observations of starspots for fully convective
stars.

\end{abstract}

\section{Introduction}

M-dwarfs are the most abundant stars in the Solar neighborhood, and among
the most magnetically active.  Often this magnetism is observed through
photometric and Doppler imaging of starspots
\citep[e.g.][]{davenport2015,barnes2015}, or monitoring of coronal or chromospheric
activity \citep[e.g.][]{hawley1996,pizzolato2003,wright2011}.  Measurements of
Zeeman broadening on magnetically sensitive atomic lines or molecular bands
suggest average surface magnetic fields reaching a few kG
\citep[e.g.][]{johnskrull1996,reiners2007}, comparable to the magnetic
field observed in sunspots \citep[e.g.][]{solankireview,borrero2011}.
Violent flares rivaling those of the Sun are commonplace
\citep[e.g.][]{hilton2010,osten2010}, calling into question the
habitability of planets in an M-dwarf system
\citep[e.g.][]{segura2010,cohen2014}.

This magnetic activity ultimately arises from dynamo action occurring
within the star.  Many dynamical processes may contribute to the
operation of the dynamo, but convection, rotation, and shear are all thought to
play particularly significant roles \citep[see e.g.][]{moffatt1978,brandenburgphysrep2005,miesch2009}.  
In this paper, we explore the possibility that fibril magnetic fields are generated by dynamo action in the interiors of low-mass stars, and rise to the surface (via magnetic buoyancy) where they might be observed.  We examine the rise of these magnetic structures via simulations within the thin flux tube approximation.  This approach has been widely used in the Solar context, but not previously applied to fully convective stars \citep[apart from the brief analysis in][]{browning2016}. In this section, we outline some observational aspects of stellar magnetism that motivate and guide our work, and review previous theoretical studies of magnetic fields in M-dwarfs.

\subsection{Magnetism Across the Tachocline Divide}

It has long been
argued that the seat of the global dynamo in solar-type stars resides in
the tachocline, a region of shear at the interface between the
differentially rotating convection zone and the stably stratified (and
rigidly rotating) interior \citep[e.g.][]{spiegel1980,
  parker1993,char1997,ossendrijver2003}.  In this classic ``interface
dynamo,'' the toroidal magnetic field is amplified and stored in the stably
stratified tachocline until becoming unstable to magnetic buoyancy
instabilities \citep[see e.g.][]{hughes1988,hughes2007,fan2009,charlrsp2010,cheung2014}. Portions of the field rise through the convection zone; some are shredded, others may be ``pumped''
downward by the convection \citep[e.g.][]{tobias2001}, but some rise to
the surface where they may be observed as active regions.  In this model,
the appearance of starspots in distinct latitudinal bands on the Sun and
some Solar-like stars \citep[e.g.][]{barnes1998} results essentially from the combined
effects of magnetic buoyancy, magnetic tension, and Coriolis forces that
influence fibril magnetic structures as they rise \citep[for a review, see e.g.][]{fan2009}.
Some recent models have adopted the view that the tachocline may not play as
crucial a role as previously believed \citep[see e.g.][and discussion below]{brandenburg2005}.  But in many of these, the differential rotation is still regarded as crucial: either as a
direct source of poloidal field from toroidal, or as a way of mitigating
the effects of small-scale dynamo action \citep{tobias2013,cattaneo2014}.  

Stars with masses $\lesssim 0.35$ M$_{\odot}$ (near the transition from
spectral types M3 to M4) are fully convective, and so lack an interface
region akin to the Solar tachocline.  But clearly stars on either side of this
``tachocline divide'' still effectively build magnetic fields -- with,
e.g., chromospheric H$\alpha$ emission, a common proxy for magnetic
activity, increasingly prevalent in the late-M spectral types
\citep[e.g.][]{hawley1996,west2004, schmidt2015}.  Further, in at
least some ways that magnetism is akin to that observed in Sun-like stars:
in particular, there is still some form of rotation-activity correlation \citep[e.g.][]{reiners2009,wright2011,reinersaj2012,west2015}.  Although such stars do not possess a tachocline, it is natural to wonder whether internal shear (i.e., differential rotation) might play a similar role, and whether any other signatures of the transition to full convection might be found in, for example, the patterns of magnetic activity visible at their surfaces.


Observations suggest that surface differential rotation is comparatively
smaller for stars of lower effective temperature
\citep[e.g.][]{reinhold2013}, similar to the trend found from mean-field
models \citep[e.g.][]{kuker2011} and from 3D dynamo simulations (discussed below).  The surface shear from the
equator to the pole of some M4 dwarfs indicates essentially solid body rotation \citep[e.g.][]{morin2008,davenport2015}.  Observations of magnetic ``spots'' at the surface of a star encode, in
principle, information about the generation of fields and their rise to the
surface.  In contrast to the preferred toroidal belts of activity observed
on the Sun, starspots on even rapidly rotating M-dwarfs have been observed at all latitudes
(though clearly with far less precision than is possible in the Solar
case) { \citep[e.g.][]{davenport2015,barnes2015}.}  It may be the case
that polar starspot caps are the result of a predominantly dipolar magnetic
field topology, as seen in some 3D global dynamo models of fully convective
stars \citep[e.g.][]{gastine2013,yadav2015aanda, yadav2015}.  However, it is also
known that rapid rotation has a tendency to deflect buoyantly rising flux
tubes poleward \citep[e.g.][]{schuessler1992,deluca1997}.  If such tubes
were generated in the interior, they would also then tend to emerge near
the poles. 

Spots constitute only a part of the surface magnetism.  While the surface
magnetic field may reach a few kG, \citet{reiners2009} report that more
than $\sim85\%$ of the magnetic flux in early-to-mid M-dwarfs is on small
scales.  Furthermore, a reduced starspot-induced light-curve variability in
mid-to-late M-dwarfs compared to earlier spectral types suggest a more
uniform distribution of starspots \citep[see
  e.g.][]{messina2003,rockenfeller2006,jackson2012}.  At some level, every star
likely possesses a unique magnetic field topology, with both large-scale
and small-scale surface fields contributing to the overall observed
magnetic field strength.  What remains clear is that dynamos in fully
convective stars are capable of producing strong magnetic activity, in some
cases without evident differential rotation.  These magnetic fields may lead to observable
starspots, which could exceed the latitude and
filling factor of spots on the Sun.

\subsection{Prior Modeling and This Work}

The generation of magnetic fields by dynamo action in fully convective stars is an intricate process, and not especially well understood.  One early suggestion was that such objects would host only a turbulent magnetic field, structured on small spatial scales \citep{durney1993}.  Within the context of mean-field theory, \citet{chabrier2006} later suggested that such stars host $\alpha^2$ dynamos, with helical motions as the source of both the poloidal and toroidal magnetic fields. 
Several authors have turned to magnetohydrodynamic (MHD) numerical simulations as a way of gauging the strength
and morphology of the magnetism, which in turn arises from the combined
influences of convection, rotation, and shear.  In the context of fully
convective stars specifically, \citet{dobler2006,browning2008,yadav2015}
have found that fields with a wide range of spatial
scales can be built by the flows.  Under the strong rotational constraints
typical of M-dwarfs, a non-trivial large-scale magnetic field component can be generated.
In the global, anelastic MHD models of \citet{browning2008} and \citet{yadav2015}, for example, the
overall magnetic field grows until it is roughly in equipartition with the
kinetic energy (corresponding to average strengths of $\sim 2-10$ kG in
different regions of the spherical domain). In both cases, the strong magnetism
acts to reduce differential rotation present in hydrodynamic progenitor
calculations.  The spatial structure of the field varies somewhat in
different models. In \citet{yadav2015} the field exhibits a strong
dipolar component (coexisting with smaller-scale features), whereas in
\citet{browning2008} the magnetic energy spectrum peaks at somewhat higher
spherical harmonic degrees (i.e., smaller spatial scales).  In the context
of Solar-like stars, simulations have likewise suggested that strong and coherent mean fields (both toroidal and poloidal)
may be built without a ``tachocline'' of shear, particularly if the overall
influence of rotation is strong enough.  In the simulations of \citet{brown2010},
\citet{brown2011}, or \citet{augustson2015}, for example,
coherent ``wreaths'' of toroidal field are built amidst the convection,
provided it is rotating rapidly enough.  Though these simulations operate
in parameter regimes far removed from those realized in actual stars, surely influencing the character of the magnetism \citep[see e.g., discussions in][]{cattaneo2009,tobias2011}, they are nonetheless suggestive of the sorts of organized fields that might be built by
convection and rotation.

These global-scale simulations are just beginning to capture some aspects
of {\sl magnetic buoyancy}, long thought to play a role in stellar dynamos.  Magnetic buoyancy has been studied extensively using both analytical theory and simulations in
localized domains \citep[see e.g.][]{parker1955,parker1975,newcomb1961,acheson1979,hughes1988,fan2009,cheung2014}.  In particular, the global simulations of a rapidly rotating convective envelope (with no tachocline) by \citet{nelson2011,nelson2013} self-consistently generate toroidal magnetic structures that rise under the combined influence of magnetic buoyancy and advection by convective flows.  In those simulations, such structures arise
essentially as the high-strength tail of an extended distribution of
field strengths: while typical field strengths are only a few kG, the
buoyant loops occur only in regions with field strengths $> 35$ kG. The
models described there are extraordinarily expensive to compute.  Indeed
it is only by reaching a particularly low level of diffusion (achieved
through the use of a dynamic Smagorinsky sub-grid-scale model) that the
buoyant loops begin to emerge naturally.


Any real stellar dynamo will produce magnetic structures with a wide range
of spatial scales and field strengths.  Even in the absence of strong
internal differential rotation, bands of coherent toroidal field may arise.
For example, the simulations of \citet{yadav2015} or \citet{browning2008} 
both yield toroidal fields that exceed 10 kG (and are greater than
the associated poloidal fields), despite the weak zonal flows.  Because the convection is comparatively weak in M-dwarfs, and rotation
is often relatively rapid, even modest angular velocity contrasts of $\Delta
\Omega/\Omega \sim 10^{-3}$ can still yield a considerable influence on the
dynamo.  Turning to the simulations of \citet{nelson2011,nelson2013} as a
qualitative guide, we might expect that pushing the simulations to
even lower values of diffusivity (and commensurately more turbulent flows) would result in toroidal fields with stronger portions, even if the
mean level of magnetic energy were not greatly changed.  The strongest of
these fields must feel the effects of magnetic buoyancy, and so begin to
rise.  

The journey of magnetic fields from their region of generation to the stellar surface is complex.  Often the flux tube model is adopted to describe the dynamic evolution of magnetic field bundles.  A rich body of work applying the flux tube model in the Solar context has provided important insight into the flux emergence process.  These simulations have been performed in both horizontal and spherical geometries, utilizing either a fully 3D MHD approach or the effectively 1D TFT approximation \citep[for a comprehensive review, see e.g.][]{fan2009}.

Thin flux tube calculations have been useful in understanding the mechanisms driving the observed properties of Solar active regions.  They have been particularly helpful in elucidating the role that Coriolis force plays in determining the latitude of emergence \citep[e.g.][]{choudhuri1989,fan1993,caligari1995}, tilt of the active region toward the equator \citep[e.g.][]{dsilva1993,caligari1995}, and morphological \citep[e.g.][]{fan1993,caligari1998} and geometrical \citep[e.g.][]{moreno1994,caligari1995} asymmetries. Unlike TFT simulations, flux tube simulations of the 3D variety can resolve the cross-section of the tube and twist of the internal magnetic field lines. These simulations also capture the back-reaction of the magnetic structures on the surrounding plasma and the possible shredding of the flux tube by convection \citep[e.g.][]{fan2003,abbett2004,jouve2009,pinto2013}.  However, due to the limited numerical resolution and relatively high imposed magnetic diffusion of such 3D models, tubes with strong super-equipartition magnetic field strength and large radii are typically required, corresponding to a total flux often larger than observed active regions on the Sun (i.e. $\gtrsim10^{23}$ Mx).  Note, though, that the radii of flux tubes in some 3D models, for instance \citet{fan2008} and \citet{jouve2009}, are only $\sim$3 times larger than in the simulations we present here (see Sec. \ref{sec:init}).


Although flux tube models cannot address the self-consistent formation of magnetic field bundles, they are nonetheless instructive.  In particular, they allow the flexibility of prescribing initial conditions both of the flux tube and the external environment to explore a variety of possible situations that may be realized in stars.  \citet{weber2011,webersolphys2013,weber2015} examine the effect Solar-like convection has on the local and global evolution of magnetic flux tubes while circumventing the problem of artificial diffusion by employing the TFT model in a hydrodynamic convection simulation.  While idealized, these simulations complement the results of both 3D MHD flux tube simulations and those of the buoyantly rising loops generated through dynamo action as in \citet{nelson2014}.  Namely, they show that both magnetic buoyancy and convection contribute to the flux emergence process, acting in concert to replicate the observed properties of Solar active regions.  Additionally, as the TFT model is a 1D code, simulations of individual flux tubes may be performed quickly on single processor machines, much faster than 3D simulations requiring millions of processor hours on massively parallel supercomputers.

Inspired by the growing number of observations of fully convective stars
and encouraged by the results obtained from previous TFT simulations, we
turn here to simulations of thin flux tubes embedded in fluid motions representative of a fully convective star.  Our aim is to investigate whether toroidal fields built in the bulk
of the convection zone could potentially give rise to the starspots
observed on fully convective M-dwarfs.  Our approach adopts a number of
simplistic assumptions: most significantly, we have assumed that the dynamo generated
magnetic field produces coherent, toroidal tubes of field.  The traditional
TFT model assumes that this magnetic field is generated by an
interface dynamo at the boundary of the radiative interior and convective
envelope.  Here, in effect we assume that a distributed dynamo is capable of building
toroidal flux tubes as well.

In Section \ref{sec:model}, we introduce our model and initial flux tube conditions.  Section \ref{sec:quiescent} describes the evolution of axisymmetric flux tubes in a quiescent interior, both initially in temperature equilibrium (Sections \ref{sec:dynamic}-\ref{sec:radheat}) and in comparison to those in mechanical equilibrium (Section \ref{sec:mecheq}).  We present the results of our TFT simulations embedded in a hydrodynamic convective flow field in Section \ref{sec:tubes_conv}, focusing on the latitude of emergence and the effect of differential rotation in Section \ref{sec:diff}, and the efficiency of magnetic pumping in Section \ref{sec:pump}.  We conclude and reflect on our results in Section \ref{sec:conclude}.

\section{Formulating the Problem}
\label{sec:model}
\subsection{Modeling Fibril Magnetic Fields}
\label{sec:TFTeq}
The dynamics of thin, isolated magnetic flux tubes can be described by invoking the thin flux tube (TFT) approximation \citep[e.g.][]{roberts1978,ferriz1989,spruit1981b}.  The TFT equations are derived from ideal MHD, operating under the assumption that all variables are constant over the cross-sectional radius $a$ of the flux tube.  Consequently, the set of equations is reduced to one spatial dimension, with all quantities represented by their values along the flux tube axis.  The equations that describe the evolution of each Lagrangian element of the 1D flux tube are as follows:
\begin{eqnarray}
\rho {d {\bf v} \over dt} & = & -2 \rho ( {\bf \Omega_0} \times {\bf v} )
-(\rho_e - \rho ) {\bf g}  + {\bf l} {\partial \over \partial s} \left ( { B^2 \over 8 \pi}
\right ) + {B^2 \over 4 \pi} {\bf k} 
\nonumber \\
& & - C_d {\rho_e | ({\bf v }-{\bf v}_e)_{\perp} |
({\bf v}-{\bf v}_e)_{\perp} \over ( \pi \Phi / B )^{1/2} }, 
\label{eq:eqn_motion}
\end{eqnarray}
\begin{eqnarray}
{d \over dt} \left ( {B \over \rho} \right )  &  = &   {B \over \rho} \left [
{\partial ({\bf v} \cdot {\bf l}) \over \partial s} - {\bf v} \cdot
{\bf k} \right ] , 
\label{eqn_cont_induc} \\
\nonumber \\
{1 \over \rho} {d \rho \over dt} & = &   {1 \over \gamma p} {dp \over dt}-\nabla_{ad}\frac{\rho}{p}T\frac{dS}{dt} ,
\label{eqn_adiab} \\
\nonumber \\
p  & = & {\rho R T \over \mu} ,
\label{eqn_state} \\
\nonumber \\
p_e & = & p + {B^2 \over 8 \pi} ,
\label{eqn_pbalance}
\end{eqnarray} 
where, ${\bf r}$, ${\bf v}$, \emph{B}, \emph{$\rho$}, \emph{p}, \emph{T}, which are functions of time \emph{t} and arc length \emph{s} measured 
along the tube, denote respectively the position, velocity, magnetic field strength, gas density, pressure, and temperature of a Lagrangian tube segment, ${\bf l} \equiv \partial {\bf r} / \partial s$ is the unit vector tangential to the flux tube, and ${\bf k} \equiv \partial^2 {\bf r} / \partial s^2 $ is the tube's curvature vector, subscript ${\perp}$ denotes the component perpendicular to the flux tube, $\Phi$ is the constant total flux of the tube, $\rho_{e}$, $p_{e}$, and $\mu$, which are functions of depth only, are the pressure, density, and mean molecular weight of the surrounding external plasma, ${\bf g}$ is the gravitational acceleration and a function of depth, ${\bf \Omega_{0}}$ is the angular velocity of the reference frame co-rotating with the star, with $\Omega_{0}$ set to the typical solar rotation rate of 2.6\e{-6} rad s$^{-1}$,  $C_d$ is the drag coefficient set to unity, $\gamma$ is the adiabatic exponent $(\partial  \ln{p}/\partial \ln{\rho})_{ad}$, $S$ is the entropy per unit mass, $\nabla_{ad}$ is the adiabatic temperature gradient $(d \ln{T}/ d \ln{p})_{ad}$ assumed to be the same as the background plasma, $R$ is the ideal gas constant, and ${\bf v}_e ({\bf r}, t)$ is an external velocity field relative to the rotating frame of reference that impacts the dynamics of the TFT through the drag force term, described in Section \ref{sec:ASH}.  The term ${\bf v}_e ({\bf r}, t)$ accounts for both the local convective flows and mean flows such as differential rotation.   

In the above equations, we do not introduce an explicit magnetic diffusion or kinematic viscosity term.  The TFT approximation preserves the frozen-in condition of the magnetic fields, proceeding with an effectively infinite magnetic Reynolds number.  The flux tube evolves passively in the external fluid, imparting no back reaction on the fluid in which it is embedded. The magnetic field of the flux tube is untwisted such that it only has a component in the $\hat{l}$ direction, and the tube is discretized with 800 uniformly spaced mesh points along its total length.  A description of the numerical methods used to solve the flux tube evolution as determined by the above set of equations is discussed in detail by \citet{fan1993}.  The fundamental simulation code is the same as in \citet{weber2015}, extended here to a fully convective interior.  Stratification and thermodynamic properties of the external field-free plasma are taken from a one-dimensional stellar structure model of a fully convective 0.3M$_{\odot}$ main-sequence star provided by Isabelle Baraffe following \citet{chabrier1997}.    

\begin{figure}
\centering
\includegraphics[scale=.42,trim=2cm 0cm 0cm 0cm]{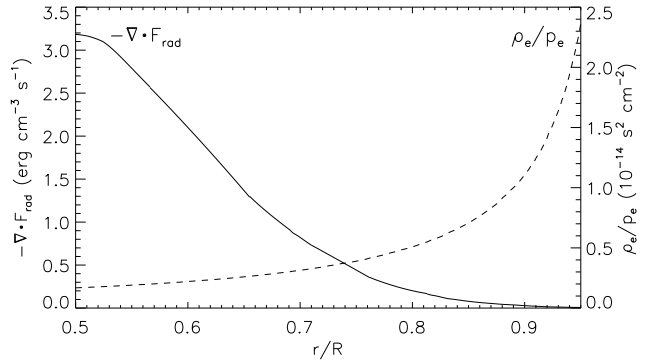}
\caption{Profiles of $-\nabla \cdot {\bf F}_{rad}$ (left axis) and $\rho_{e}/p_{e}$ (right axis) across our domain of interest from 0.5-0.95R.  Radiative heating decreases across the layer while $\rho_{e}/p_{e}$, which is directly proportional to the buoyancy of the flux tube, increases toward the surface.}
\label{fig:divfrad}
\end{figure}

The last term on the right-hand side of Eq. \ref{eqn_adiab} contains the rate of heat input per unit volume to the flux tube plasma, which can be reduced to two dominant terms \citep[see][]{fan1996,weber2015}:
 \begin{equation}
 \rho T \frac{dS}{dt} \approx -\nabla \cdot {\bf F}_{rad} - \kappa_{e} \frac{\alpha_{1}^{2}}{a^{2}} (T-T_{e}),
 \label{eqn:rad}
 \end{equation} 
where ${\bf F}_{rad}$ is the radiative energy flux vector, $\kappa_{e}$ is the radiative conductivity, $\alpha_{1}$ is the first zero of the Bessel function $J_{0}(x)$, and $a=(\Phi/\pi B)^{1/2}$ is the cross-sectional radius of the flux tube.  Thermodynamic values ${F}_{rad}$, $\kappa_{e}$, and $T_{e}$ are taken from the stellar structure model.  The first term on the right-hand side of Eq. \ref{eqn:rad} is the divergence of radiative heat flux in the background plasma (see Fig. \ref{fig:divfrad}).  The second term represents a radiative diffusion across the flux tube due to temperature differences between the flux tube and external plasma.  In Section \ref{sec:radheat}, we will discuss the implications of including radiative heating in the flux tube model.  Adiabatic evolution occurs when $dS/dt=0$.    

\subsection{Flux Tube Initial Conditions}
\label{sec:init}
In most of what follows, we operate under the simplifying assumption that large-scale, turbulent convective motions have built perfectly toroidal flux tubes initially in thermal equilibrium (hereafter TEQ) with the background fluid.  The condition of TEQ deviates from the state of neutral buoyancy ($\rho_{e}=\rho$) often used in TFT simulations for stars with convectively stable radiative interiors \citep[e.g.][]{caligari1998,granzer2000,holzwarth2001,weber2011}.  Within the long-favored Solar interface dynamo paradigm, it is assumed that toroidal magnetic fields are amplified and stored in the stably stratified convective overshoot/tachocline region.  Even if magnetic flux tubes were built in this region in TEQ, the subadiabatic stratification of the radiative interior would cool the tube as it rose through the region, eventually achieving a state of neutral buoyancy \citep[e.g.][]{moreno1992}.  For stars with fully convective interiors though, it is difficult to imagine a scenario where a flux tube may achieve neutral buoyancy.  To that end, we assume that our flux tubes are in TEQ with the surrounding fluid, inducing a density deficit,
\begin{equation}
\rho_{e}-\rho = \rho_{e}\frac{B_{0}^{2}}{8 \pi p_{e}},
\label{eq:tempeq}
\end{equation} 
rendering the tube initially buoyant.  The ratio of $\rho_{e}/p_{e}$ increases approaching the stellar surface (see Fig. \ref{fig:divfrad}).  Therefore, flux tube generated in shallower layers will be more buoyant than those generated deeper.

For the purposes of this paper, we are interested in studying magnetic flux tubes that may give rise to starspots.  Simulations suggest that dynamo action in fully convective stars can achieve kG-strength magnetic fields, roughly in equipartition with convective motions, without a tachocline region \citep[e.g.][]{dobler2006,browning2008,yadav2015aanda}.  To rise toward the surface without experiencing significant downward pumping by convection, the magnetic field strength must exceed a critical level such that the buoyancy of the flux tube dominates the downward drag force from the convective flows.  An estimate of this critical field strength is a few times the equipartition value, given by $B_{c}\ge(H_{p}/a)^{1/2} B_{eq}$ \citep[see e.g.][]{fan2003}, where $B_{eq}=\sqrt{4 \pi \rho}v_{c}$.  For perspective, the equipartition field strength varies with depth, but is typically between 2 and 10 kG.  Below $B_{c}$, flux tubes may become severely distorted by convection, unable to retain cohesion.  It is therefore likely that only the most extreme, high field strength events in the dynamo-generated magnetic field pdf become buoyant enough to make the journey toward the surface, as suggested by \citet{nelson2014}.        

There are also limits on how large the dynamo-generated magnetic field can become.  To explain the observed inflation of some low-mass stellar radii compared to 1D structure models, the presence of strong $1-100$ MG fields have been suggested \citep[e.g.][]{mullan2001,feiden2014}.  However, for such extreme magnetic field values, \citet{browning2016} have shown that large-scale flux tubes (i.e. larger cross-sectional radius $a$) will rise to the surface faster than they can plausibly be generated by large-scale convective eddies.  Furthermore, the dissipative ohmic heating associated with a small-scale MG field (i.e. smaller $a$) may exceed the luminosity of the star.   

In light of these magnetic field constraints, we have chosen to investigate flux tubes of $B_{0}=30-200$ kG to capture both the lower and higher end of the magnetic field strength range.  Within this bound, we present simulations at six discreet values of $B_{0}=$ 30, 40, 60, 80, 100, and 200 kG.  As the flux tube traverses the convection zone, the magnetic field strength decreases proportionally to the density profile $B \propto \rho^{\alpha}$ \citep[see e.g.][]{fan2001,cheung2010,pinto2013}.  This relation is derived on the assumption that both the mass and magnetic flux of the tube are conserved.  An exponent $\alpha=1$ is expected for tubes that expand as they rise without changing length, while smaller values are expected if the tube is substantially stretched.  Assuming $\alpha \sim 1$, such initial fields imply magnetic field strengths of about an order of magnitude less at the simulation upper boundary (0.95R), and two orders of magnitude at 0.99R, giving rise to tubes at this depth of 300 - 2000 G.  If the flux tubes survive the remaining 0.01R without significant shredding by convective motions or depletion of magnetic field, the legs of the loop will intersect with the photosphere.  Radiative cooling of the plasma inside the tube in combination with the strong super-adiabatic gradient could initiate the process of `convective collapse'.  This intensifies the flux tube to higher magnetic field strengths, giving rise to cooler regions marked by suppressed convection, such as starspots, pores, and faculae \citep[see e.g.][]{parker1978,spruit1979collapse,spruitzweibel1979}.  

Flux tubes are initiated at two different depths, 0.5R and 0.75R, in order to sample the differing convective flow pattern and differential rotation structure with depth.  At such depths, the magnetic field strength in equipartition with the rms radial downflows is $\sim$8 kG and $\sim$6 kG, respectively.  Therefore, at 0.5R, the range of $B_{0}$ we consider is $\sim$$4-25 B_{eq}$ and $\sim$$1-5.5 B_{c}$.  At a depth of 0.75R, $B_{0}$ is $\sim$$5-33 B_{eq}$ and $\sim$$1.5-10 B_{c}$.  Hence, fields significantly weaker than those considered here would likely be highly susceptible to downward magnetic pumping.  The initial latitude is also varied from $0^{\circ}-60^{\circ}$ in both hemispheres, with $1^{\circ}$ intervals between $0^{\circ}-15^{\circ}$, and $5^{\circ}$ intervals from $15^{\circ}-60^{\circ}$.


A constraint of the TFT approximation requires that $a/H_{p} \le 0.1$ in the region where the flux tube is initiated \citep[e.g.][]{fan1993}.  The pressure scale height $H_{p}$$\sim$1.7\e{9} cm at 0.75R, indicating that the maximum allowable initial cross-sectional radius of the flux tube at this height is 1.7\e{8} cm.  Assuming that the total flux of the tube remains constant, given by $\Phi=B \pi a^{2}$, the magnetic flux ranges from 2.72\e{21} Mx for 30 kG tubes to 1.82\e{22} Mx for 200 kG tubes.  This range of magnetic flux is typical of active regions found on the Sun \citep[e.g.][]{zwaan1987}.  However, the total unsigned magnetic flux of some active G, K, and M-dwarfs can exceed that of Solar disk averages by upwards of 3 orders of magnitude \citep{pevtsov2003flux}.  The contribution to the total magnetic flux from starspots is a function of the area, magnetic field strength, and number of individual starspot regions.  The larger total surface flux for later spectral types could be explained in part by starspots of roughly the same magnetic field strength as the Sun covering a larger stellar surface area, or spots of stronger magnetic fields covering a similar stellar surface area as observed on the Sun.  If fully convective stars indeed have a greater spot coverage than Solar-type stars, then the former scenario may be more likely \citep[for other observational constraints on this point, see e.g.][]{reinersetal2009}.  This could be achieved by the flux tubes we model if many individual tubes appear at the surface distributed randomly on the star, or if bunches of individual flux tubes rise to create extended dark spots of suppressed convection.       

To keep our investigation here computationally tractable, we only perform simulations where $a=$1.7\e{8} cm.  This value is rather arbitrary, but again, is the maximum allowable $a$ under the TFT approximation at 0.75R.  As the drag force acting on the flux tube is proportional to $a$, this ensures that all flux tubes initiated at the same depth will experience a drag force of roughly the same magnitude early in their evolution.  Furthermore, the cross-sectional radius $a$ ought to remain small across the domain, ideally less than a few times the pressure scale height $H_{p}$.  As the flux tube nears the surface, the cross-sectional radius expands quickly due to the more rapid decrease of pressure and density of the external plasma.  As a result, we stop our simulations once the flux tube has reached $0.95R$, operating under the assumption that the rise time and trajectory through the remaining $0.05R$ is negligible compared to the total rise. 

In Sections \ref{sec:dynamic} and \ref{sec:radheat}, we study, respectively, how the evolution of flux tubes rising in a quiescent convection zone respond to the choice of initial internal rotation rate $v_{\phi 0}$ and the addition of radiative heating.  To facilitate comparison to previous TFT simulations, we briefly discuss the difference between flux tubes initiated in mechanical equilibrium as opposed to thermal equilibrium in Section \ref{sec:mecheq}.  The results of our flux tube simulations incorporating a convective flow field are discussed in Section \ref{sec:tubes_conv}.

\subsection{Convective Velocity Field}
\label{sec:ASH}  

To capture the influence of global scale convection on flux emergence in a fully convective star, a convective flow field computed separately from the TFT simulations is incorporated through the aerodynamic drag force acting on each flux tube segment (last term in Eq. \ref{eq:eqn_motion}).  We utilize the Anelastic Spherical Harmonic (ASH) code, which solves the 3D magnetohydrodynamic (MHD) equations within the anelastic approximation.  The progenitor case of the velocity field we use here is identical to the hydrodynamic simulation of Case C in \citet{browning2008}.  Representative of fluid motions in fully convective stars, this ASH simulation captures giant-cell convection and the associated mean flows such as differential rotation in a rotating spherical domain spanning from 0.10R to 0.97R, with R the total stellar radius of 2.013\e{10} cm.  ASH is a pseudo-spectral code, here resolved by a grid of 127 points in $r$, 256 points in $\theta$, and 512 points in $\phi$.      

\begin{figure}
\includegraphics[scale=1,clip=true,trim=4.5cm 0cm 4.5cm 0cm]{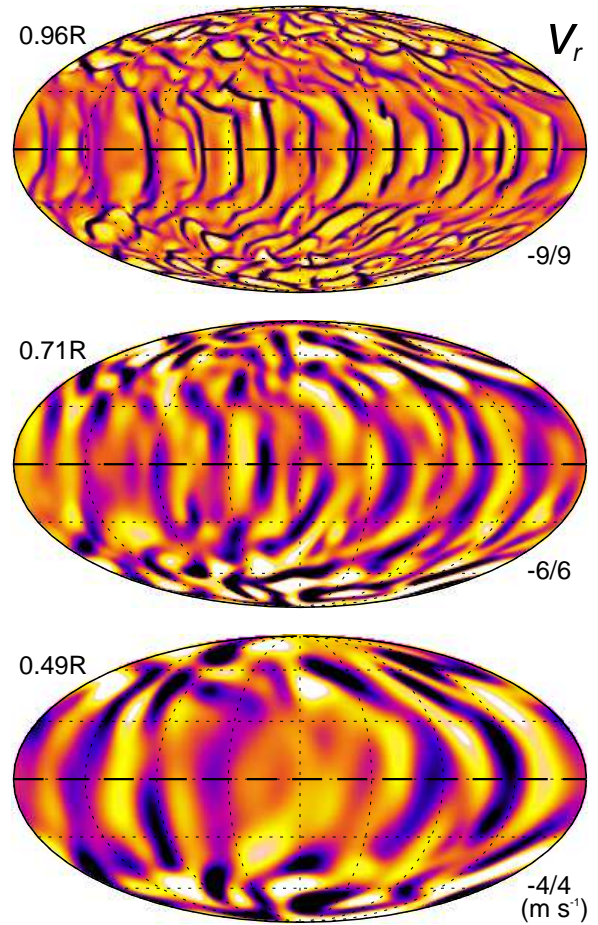}
\caption{Radial velocity $v_{r}$ snapshots on spherical surfaces at three depths taken at the same instant, shown in Mollweide projection. Upflows are rendered in red tones, and downflows in blue; saturation values are indicated.  Flows are stronger and on smaller spatial scales near the surface than they are at depth.}
\label{fig:ash_vr}
\end{figure}

\begin{figure}
\centering
\includegraphics[scale=.54,trim=.2cm 0cm 1.8cm 0cm]{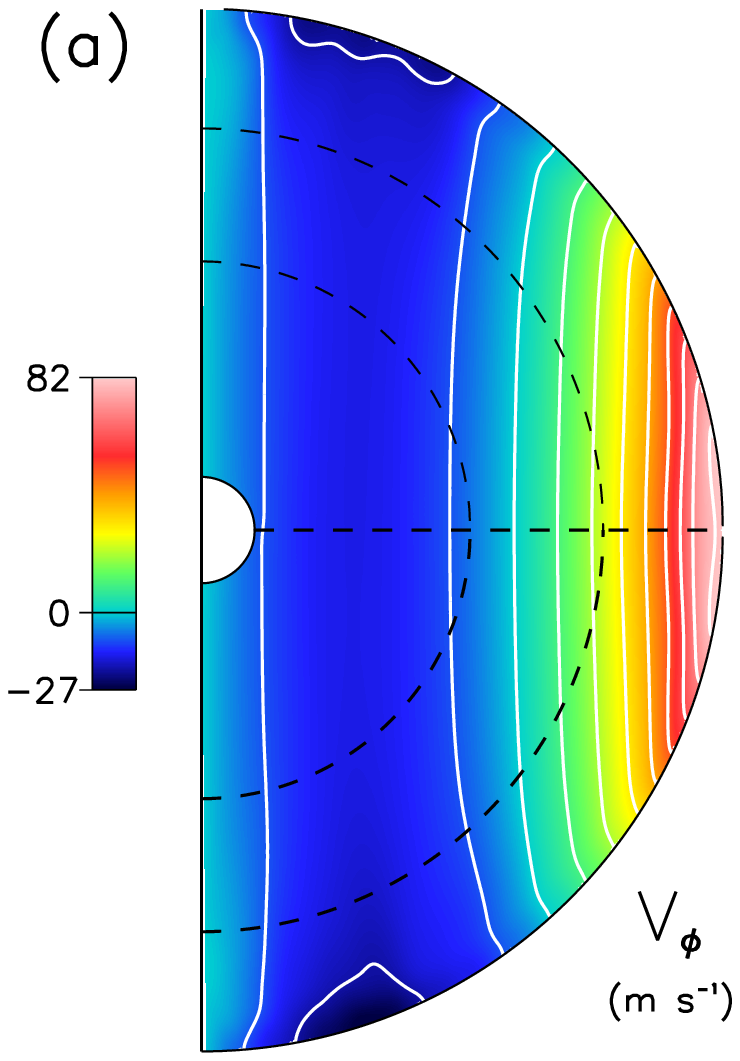}
\includegraphics[scale=.54]{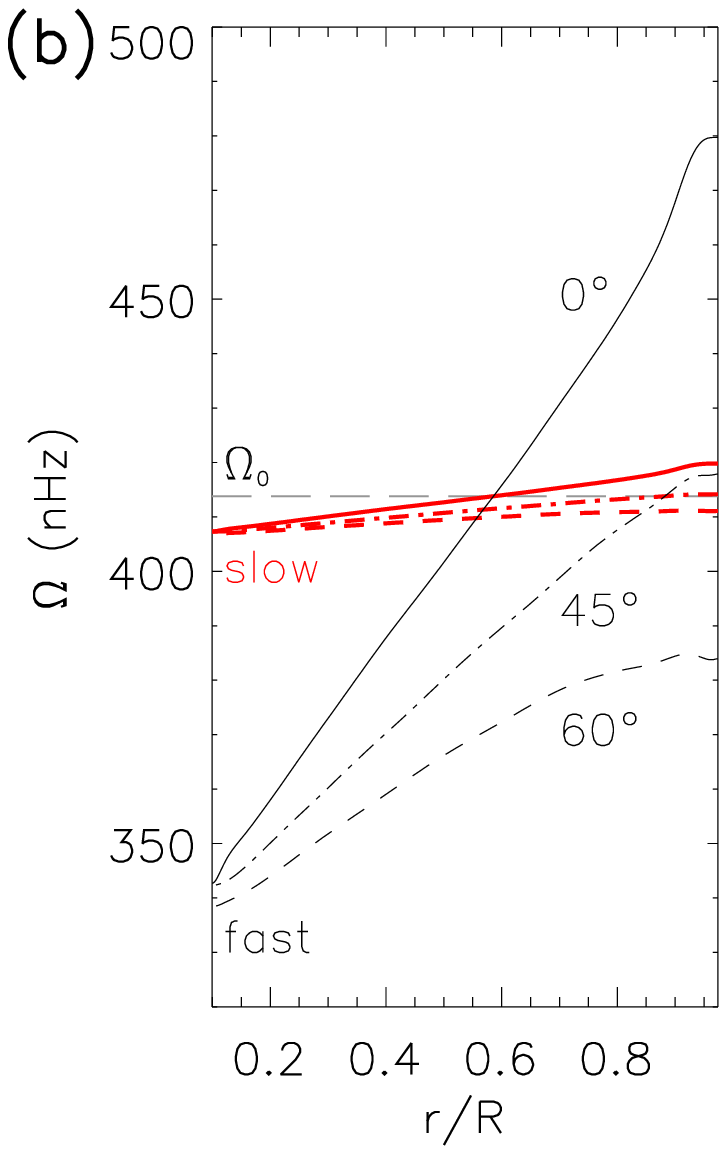}

\caption{ {\bf (a)} Meridional plot of the longitudinal velocity $\hat{v}_{\phi}$ for the \emph{fast} differential rotation profile, averaged over $\sim460$ days with contour intervals every 10 m s$^{-1}$ around zero relative to the rotating frame.  Dashed lines are at radii of 0.5R and 0.75R. {\bf (b)} Angular velocity $\hat{\Omega}$ averaged over the same time interval as a function of radius along indicated latitudinal cuts for the \emph{fast} ($\Delta \Omega/\Omega_{0}\sim 22\%$) differential rotation profile, and the \emph{slow} ($\Delta \Omega/\Omega_{0}\sim 2\%$) differential rotation profile approximated using Equation \ref{eq:approx}.} 
\label{fig:diffrotprof}
\end{figure}

 An instantaneous view of the radial velocity field at three different radii is shown in Figure \ref{fig:ash_vr}.  Typical of stratified convection, broad upflows coexist with narrower downflows.  There is a hierarchy of convective structures, with smaller downflow plumes at larger radii merging to form broader flows as they descend.  Low latitude downflow lanes have a tendency to align with the rotation axis, while the distribution is more isotropic near the poles.  The differential rotation established in the hydrodynamic simulation exhibits longitudinal velocity contours nearly parallel to the rotation axis, shown in Figure \ref{fig:diffrotprof}a, in keeping with the Taylor-Proudman constraint.  The angular velocity contrast at the surface between the equator and $60^{\circ}$ is $\Delta \Omega/\Omega_{0}$$\sim$$22\%$, comparable to the Solar angular velocity contrast of $\Delta \Omega/\Omega_{0}$$\sim$$25\%$.  When magnetism is included in the ASH simulations of \citet{browning2008}, it is found that the differential rotation is quenched, with an angular velocity contrast of $\Delta \Omega/\Omega_{0}$$\sim$$2\%$.  This result is in keeping with other 3D MHD simulations of low mass stars \citep[e.g.][]{yadav2015aanda} and is likewise in agreement with observations \citep[e.g.][]{morin2008,davenport2015}.  
   
It is likely that the strong magnetic field strengths $B_{0} > B_{eq}$ we utilize for our simulations could quench the differential rotation, reducing the angular velocity contrast to nearly that of a solid body.  For the purposes of this paper, we wish to examine how the evolution of flux tubes may change when subjected to a convection velocity field with varying degrees of angular velocity contrast.  Rather than perform multiple simulations, we retain only the time-varying radial and latitudinal components of the original hydrodynamic ASH simulation.  The azimuthal velocity field is then averaged over time and longitude, denoted by $\hat{v}_{\phi}$, and the following equation is applied to obtain a new differential rotation profile with the desired angular velocity contrast:
\begin{eqnarray}
\hat{\Omega}_{new}(\theta,r) & = & \hat{\Omega}_{orig}(\theta,r) 
\nonumber \\
& & - \bigg[ (\hat{\Omega}_{orig}(\theta,r)-\Omega_{0})*\bigg(1-
\frac{\frac{\Delta\Omega}{\Omega_{0}}|_{new}}{\frac{\Delta\Omega}{\Omega_{0}}|_{orig}}\bigg) \bigg],
\label{eq:approx}
\end{eqnarray}  
where $(\Delta \Omega/\Omega_{0})|_{new}$ and $(\Delta \Omega/\Omega_{0})|_{orig}$ are the angular velocity contrast of the new and original differential rotation profile, respectively.  This formulation centers the reduction/increase in the differential rotation profile around $\Omega_{0}$ so that it will approach solid body rotation at all depths and latitudes as $\Delta \Omega/\Omega_{0}$ is reduced.  For each timestep of the TFT simulation, a temporal and three-dimensional spatial interpolation is performed to extract the flow velocity components at the location ${\bf r}$ of each mesh point along the tube.  A two-dimensional spatial interpolation in $r$ and $\theta$ is performed on $\hat{v}_{\phi}$ to provide the longitudinal flow at each flux tube mesh point.  

Performing simulations in this way ensures that the only difference introduced comes from the applied differential rotation profile.  We recognize that a more straight-forward approach would be to use full 3D velocity fields from multiple simulations exhibiting varying degrees of angular velocity contrast.  However, the approach taken here allows for more direct comparison between TFT simulations, removing any effects that may arise because of stochastic variations in the radial $v_{r}$ and latitudinal $v_{\theta}$ velocity fields.  Figure \ref{fig:diffrotprof}b shows the angular velocity $\hat{\Omega}$ (nHz) from the original ASH hydrodynamic case as a function of radius for latitudinal cuts at $0^{\circ}$, $45^{\circ}$, and $60^{\circ}$.  Shown on the same plot is the angular velocity $\hat{\Omega}$ approximated using Equation \ref{eq:approx} for a contrast of $\Delta \Omega/\Omega_{0}$$\sim$$2\%$.  This simplistic approach creates a differential rotation profile very similar to Case Cm in \citet{browning2008} with the same angular velocity contrast of $\sim2\%$, where the presence of equipartition-strength magnetic fields quenches the differential rotation.  The presence of magnetic fields in Case Cm does affect the distribution of angular momentum.  However, we note that the amplitudes of $v_{r}$ shown in Figure \ref{fig:ash_vr} are commensurate with Case Cm in \citet{browning2008}.  Furthermore, both the hydrodynamic simulation we use here and Case Cm exhibit a similar pattern of convective cells, including a hierarchy and alignment of convective structures with the rotation axis and isotropic cells near the poles.          

In order to sample different intervals of the time-varying velocity field, we perform three ensemble simulations.  Flux tubes in each ensemble are initialized at the same moment and are advected by the exact same time-varying flow field, but evolve independently of each other.  Each ensemble is then comprised of 1176 flux tubes, one tube for each of the possible combinations of $B_{0}$, $\theta_{0}$, $r_{0}$, and applied differential rotation profile.  This equates to a total number of 3528 flux tubes analyzed in this study that evolve with the effects of convection.  The initialization times for each of the three different ensembles are arbitrary, but are at least separated by $\sim$200 days, similar to a convective turnover time in the mid-convection zone.  In Section \ref{sec:tubes_conv}, we will compare the difference between the two differential rotation profiles shown in Figure \ref{fig:diffrotprof} on flux tube evolution.  We will often refer to the two profiles as \emph{fast} (f) and \emph{slow} (s), corresponding to angular velocity contrasts $\Delta \Omega/\Omega_{0}$ of $\sim$22$\%$ and $\sim$2$\%$, respectively. 

\begin{table}
\begin{center}
\begin{tabular}{cc}
\hline
Case & Parameters \\ 
\hline
T0 & TEQ, $v_{\phi 0}=0$, Rad. Heat.\\
TL & TEQ, $v_{\phi 0}=v_{\phi \ell}$, Rad. Heat. \\
ATL & TEQ, $v_{\phi 0}=v_{\phi \ell}$, Adiabatic \\
THE & TEQ, $v_{\phi 0}=v_{\phi he}$, Rad. Heat. \\
M & MEQ, Rad. Heat. \\
\hline
f & Fast Diff. Rot., $\Delta \Omega/\Omega_{0}\sim22 \%$ \\
s & Slow Diff. Rot., $\Delta \Omega/\Omega_{0}\sim2 \%$\\
C & indicates convective field \\
\hline
\end{tabular}
\caption{Flux tube simulation parameters.  Those in TEQ have a density deficit following Eq. \ref{eq:tempeq}, with an internal azimuthal speed (1) $v_{\phi 0}=0$ co-rotating with $\Omega_{0}$, (2) $v_{\phi 0}=v_{\phi \ell}$ co-rotating with the local longitudinal velocity $\hat{v}_{\phi}$ corresponding to either the \emph{fast} or \emph{slow} differential rotation profile, or (3) $v_{\phi 0}=v_{\phi he}$, the azimuthal velocity required for the flux tube to be in horizontal force equilibrium following Eq. \ref{eq:vphi}.  Those in MEQ have a neutral buoyancy and a prograde $v_{\phi 0}$ following \citet{moreno1992}.  Flux tubes evolve either with radiative heating following Eq. \ref{eqn:rad} or adiabatically such that $dS/dt=0$.  The presence of an applied velocity field (see Sec. \ref{sec:ASH}) is represented by C.}
\label{tbl:parameters}
\end{center}
\end{table}

For simplicity in referring to a set of simulations with particular initial conditions, we have a adopted a naming scheme given in Table \ref{tbl:parameters}.  For example, the Case ATLf simulations discussed briefly in Section \ref{sec:radheat} refer to flux tubes that evolve adiabatically (A), are initially in thermal equilibrium (T), and have an internal azimuthal speed $v_{\phi 0}$ corresponding to the local longitudinal velocity $\hat{v}_{\phi}$ of the fast differential rotation profile (Lf).  The Case TLsC simulations discussed in Section \ref{sec:tubes_conv} correspond to flux tubes that evolve with the influence of radiative heating, where the tube is initially in thermal equilibrium (T) and co-rotating with the slow differential rotation profile (Ls).  The application of the suffix C indicates the presence of time-varying convective flows (C), where the applied longitudinal velocity profile $\hat{v}_{\phi}$ always corresponds to either the slow or fast profile as indicated.


\section{Flux Tubes in a Quiescent Convective Interior}
\label{sec:quiescent}
\subsection{Dynamic Evolution: Toward Horizontal Force Balance}
\label{sec:dynamic}

\begin{figure}

\centering
\includegraphics[scale=.52,clip=true,trim=1cm 0cm 1cm 0cm]{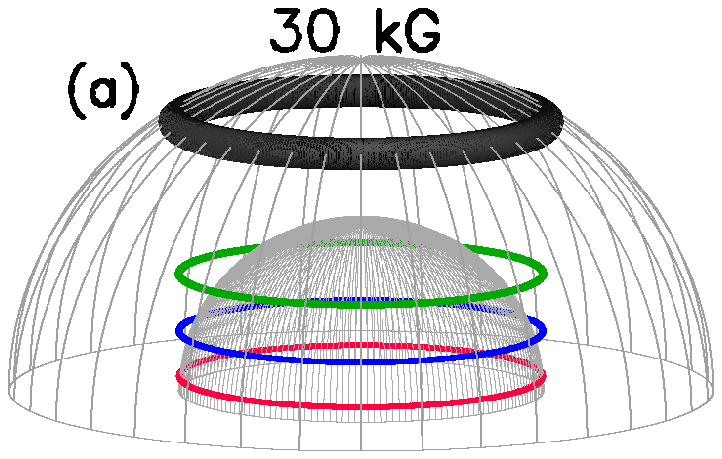}
\includegraphics[scale=.52,clip=true,trim=1cm 0cm 1cm 0cm]{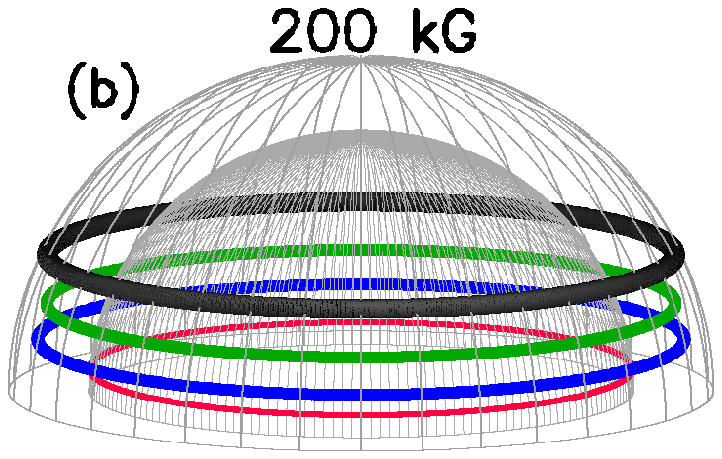} 



\caption{Time evolution of Case T0 flux tubes with $\theta_{0}=5^{\circ}$.  Inner and outer mesh spheres represent surfaces of constant radius at $r_{0}$ and 0.95R.  Flux tubes are shown at four instances: initial position (red), 50$\%$ in time through the total rise (blue), 75$\%$ (green), and once reaching the simulation upper boundary (black).   A 3D extent is applied to the tube according to the local cross-sectional radius.  Evolution of the flux tube is axisymmetric, with the trajectory largely parallel to the rotation axis.}

\vspace{.01 \textwidth}

\label{fig:no_conv}
\end{figure}

Before we examine the results from flux tube simulations allowed to evolve in a convective flow field, it is instructive to first study how axisymmetric flux tubes evolve in the quiescent interior of a fully convective star.  Figure \ref{fig:no_conv} depicts the rise of two low latitude, Case T0 flux tubes with initial magnetic field strengths and depths of ({\bf a}) $B_{0}=30$ kG, $r_{0}=0.5$R, and ({\bf b}) $B_{0}=200$ kG, $r_{0}=0.75$R.  The most striking feature is the parallel motion of the flux tube to the rotation axis.  

There are four main forces that govern flux tube evolution: buoyancy, magnetic tension, aerodynamic drag, and the Coriolis force.  The initial condition of TEQ renders the flux tube buoyant.  An inward directed (toward rotation axis) magnetic tension (curvature) force $F_{T}$ partially balances the horizontal component of the radially directed buoyancy force $F_{B}$.  The comparative magnitude of these two forces varies with depth $r$ and latitude $\theta$.  The initial ratio of the horizontal components of the buoyancy force to the magnetic tension force is given by:
\begin{equation}
\frac{F_{B}}{F_{T}}=(\rho_{e}-\rho)g \cos{\theta} \bigg[ \frac{B^{2}}{4 \pi} {\bf k}\bigg]^{-1} = \frac{1}{2 H_{p}} r_{0} \cos^{2}{\theta_{0}},
\label{eq:fbft_ratio}
\end{equation}   
where $\theta$ is the latitude, and we have used Equation \ref{eq:tempeq} and the fact that the curvature vector ${\bf k}$ has a magnitude equal to the inverse of the distance from the rotation axis.  Note that this ratio is independent of the magnetic field strength or magnetic flux, and is largest for low latitude flux tubes in shallower layers of the convection zone.  At depths of 0.5R and 0.75R for an initial latitude of $5^{\circ}$ as in Figure \ref{fig:no_conv}, the ratio $F_{B}/F_{T}$ is $\sim$1.5 and $\sim$4.4 respectively.  For comparison, a flux tube of the same $B_{0}$ at 0.5R in our 0.3M$_{\odot}$ star has a magnetic tension force $\sim$5 times larger than at the base of the solar convection zone.   

\begin{figure*}
\centering
\plotone{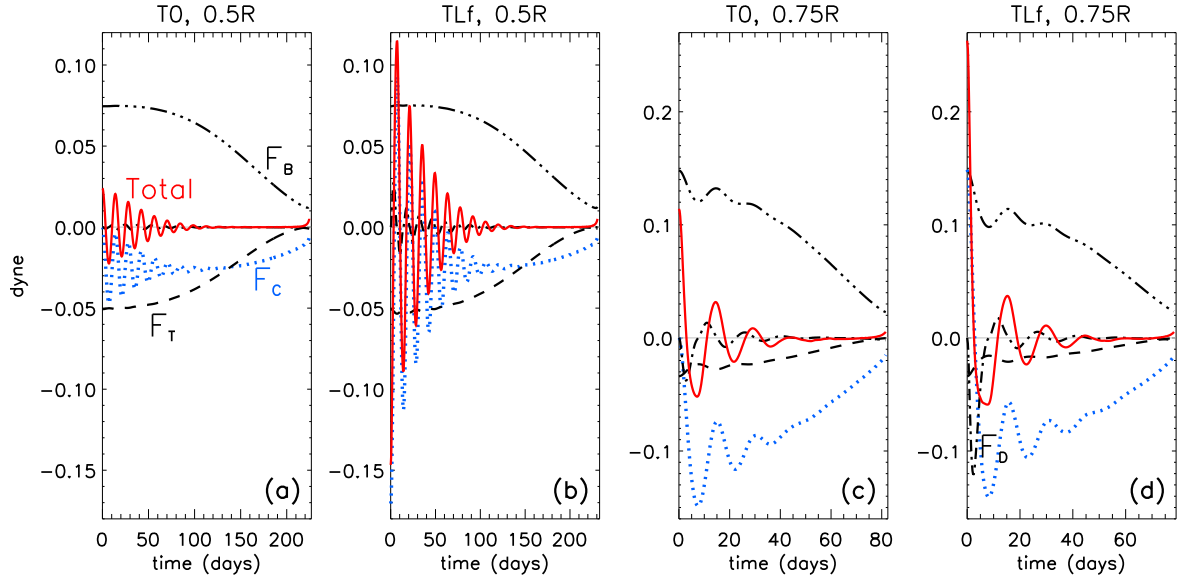}
\caption{Time evolution of the horizontal force balance for $B_{0}=80$ kG, $\theta_{0}=5^{\circ}$ flux tubes.  The sum of the horizontal forces oscillates around zero until achieving equilibrium.  Thereafter the trajectory turns mostly parallel to the rotation axis. The force components are labeled as the Total, buoyancy (F$_{B}$), Coriolis force (F$_{C}$), tension (F$_{T}$), and drag (F$_{D}$). Flux tubes in panels {\bf (c)} and {\bf(d)} are also depicted in Figure \ref{fig:choud_plot}b.}
\label{fig:horizbalance_t}
\end{figure*}

\begin{figure}
\centering
\includegraphics[scale=.92,clip=true,trim=0cm 0cm 0cm 1cm]{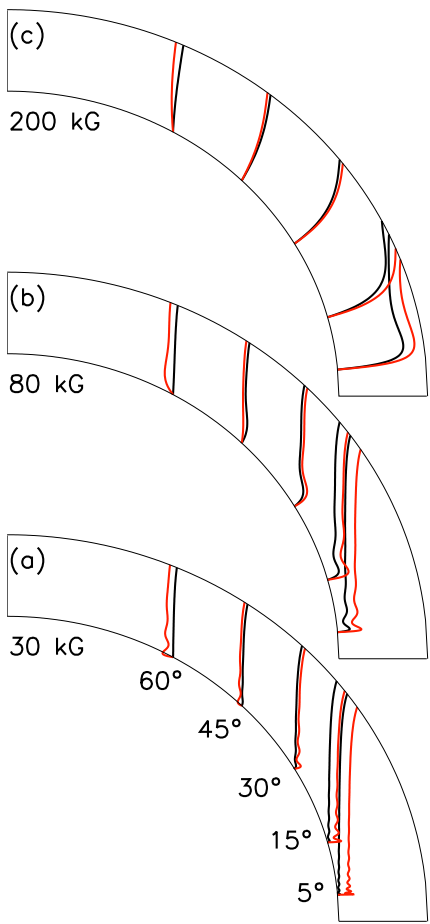}
\caption{Trajectories of flux tubes initiated at 0.75R, showing both Case T0 (black) and Case TLf (red).  A strong prograde $v_{\phi0}$ flow inside the Case TLf flux tubes can cause (via induced Coriolis forces) the tubes to move horizontally outward into shallower layers at lower latitudes, and deeper layers at higher latitudes due to a retrograde flow.}
\label{fig:choud_plot}
\end{figure}

The initial force balance of the flux tube is also sensitive to the Coriolis force.  We consider two plausible scenarios for the initial longitudinal velocity within the tube.  In one, the plasma inside the flux tube is co-rotating with the star such that $v_{\phi 0}=0$ relative to the rotating reference frame (Case T0).  In the other, the toroidal flux tube is built co-rotating with the local differential rotation.  In panels \ref{fig:horizbalance_t}a and \ref{fig:horizbalance_t}c, we show the horizontal force balance for flux tubes initially co-rotating with the star (Case T0), with panels \ref{fig:horizbalance_t}b and \ref{fig:horizbalance_t}d showing the same quantities for flux tubes given an initial $v_{\phi 0}$ corresponding to the fast differential rotation profile (Case TLf).

Figure \ref{fig:horizbalance_t} indicates that the flux tubes tend to evolve toward a state of horizontal force equilibrium.  Upon approaching horizontal equilibrium, the motion of the axisymmetric flux ring turns largely parallel to the rotation axis due to the unbalanced poleward components of the magnetic tension and Coriolis force and the vertical component of the buoyancy force.  Figure \ref{fig:choud_plot} shows the trajectories of some Case T0 and TLf flux tubes initiated at 0.75R, again illustrating the motion parallel to the rotation axis.  Similar to the axisymmetric rising flux rings of \citet{choud1987} initiated in TEQ, we find that damped oscillations, particularly in the radial and horizontal directions, can take place as the tube rises before the sum of the forces come into balance in the horizontal plane.  In the next few paragraphs, we assess in more detail the dynamics depicted in Figures \ref{fig:horizbalance_t} and \ref{fig:choud_plot}.  

Near the equator and at shallower depths, the ratio given in Equation \ref{eq:fbft_ratio} is greater than unity.  Assuming the flux tube initially rotates at the same rate as the star (Case T0), it will immediately move outward (away from rotation axis) due to the greater buoyancy force compared to tension.  The outward motion at 0.75R is much more pronounced than at 0.5R because of the greater buoyancy force there.  We do not include flux tubes initiated at 0.5R in Figure \ref{fig:choud_plot} because the horizontal oscillations are much smaller in amplitude, and the tubes deviate little from parallel motion.  This is a result of the smaller ratio of $F_{B}/F_{T}$. Conservation of angular momentum implies that a retrograde plasma flow is established inside the flux tube as it moves outward, inducing an inward directed Coriolis force.  As the tube evolves, an equilibrium of forces is established in the horizontal plane.  Once this occurs, the motion of the flux tube ceases in the horizontal direction, rising parallel to the rotation axis.  For the flux tubes shown in Figure \ref{fig:horizbalance_t}, only near our simulation upper boundary does the horizontal velocity again increase slightly in response to a diminished Coriolis force compared to the buoyancy force in the horizontal plane.  If the ratio of $F_{B}/F_{T}<1$ because the tube is in deeper layers and/or the distance from the rotation axis is small, the tube will move initially inward toward the rotation axis, inducing a prograde flow inside the tube.  It will continue to move inward until the forces roughly balance and motion parallel to the rotation axis commences.  Within the parameter space we study, such a scenario is realized for tubes initiated with latitudes $\ge35^{\circ}$ at 0.5R.  While the ratio of $F_{B}/F_{T}$ is independent of $B_{0}$, the difference between $F_{B}$ and $F_{T}$ will increase with increasing $B_{0}$, and will subsequently alter the depth in the convection zone at which the horizontal forces equilibrate and the trajectory turns poleward.     

\citet{choud1987} calculate a 2$\Omega_{0}$ frequency of oscillation for uniformly buoyant, axisymmetric flux tubes, reminiscent of inertial oscillations in a rotating fluid.  This frequency corresponds to a period of 14 days for our simulations, agreeing roughly with the oscillation periods in Figure \ref{fig:horizbalance_t}.  The greater the excess of the initial outward forces to the inward forces, the larger the oscillation amplitude and nearer the surface the flux tube moves horizontally before executing horizontal oscillations.  At 0.75R, the Case TLf flux tubes move outward into shallower layers as compared to the Case T0 flux tubes before the trajectory turns parallel to the rotation axis (see Fig. \ref{fig:choud_plot}).  This is a result of the large $v_{\phi 0}$ of the Case TLf tubes arising from the assumed co-rotation with the fast prograde differential rotation. The flux tube must move further outward, eventually establishing a retrograde flow inside the tube before the inward and outward forces come into balance.  For example, the Case T0 flux tube described in Figure \ref{fig:horizbalance_t}c and also shown in Figure \ref{fig:choud_plot}b is brought $\sim$0.03R outward from 0.75R before oscillations are initiated.  The Case TLf flux tube described in Figures \ref{fig:horizbalance_t}d and \ref{fig:choud_plot}b is brought out further to $\sim$0.06R from 0.75R before oscillations begin.  This faster motion also generates a substantial drag force opposite the direction of motion within the first $\sim$10 days (see Fig. \ref{fig:horizbalance_t}d).  Even if $F_{B}/F_{T} > 1$, some flux tubes can execute an inward trajectory before moving poleward due to a retrograde $v_{\phi 0}$ assumed from co-rotation with the differential rotation.  Such a scenario occurs for tubes initiated at 0.5R near the equatorial region (see Fig. \ref{fig:horizbalance_t}b) and also for higher latitudes at 0.75R (see Fig. \ref{fig:choud_plot}).  In summary, the difference in behavior between the Case T0 and Case TLf (or TLs) flux tubes in the quiescent convective interior is solely due to the prescribed $v_{\phi 0}$ inside the tube and the subsequent force balance established.  This study aids in our description of flux tube evolution in later sections.

The archetypal notion of rising \emph{$\Omega$-shaped loops} often discussed in the context of Solar magnetic flux emergence \citep[e.g., see review by][]{fan2009} is not realized in simulations described in this section.  The condition of TEQ means the tube will never initially be in a state of perfect force balance, and will drift away uniformly from its initial position.  This can be partially mitigated in stars with stably stratified interiors by anchoring portions of the already buoyant tube in the subadiabatic overshoot region \citep[see][]{fan1993,caligari1998}.  As we will show in Section \ref{sec:tubes_conv}, modulation of the flux tube by radial convective motions helps to pin portions of the tube to deeper layers.  Buoyantly rising loops may escape toward the surface between downdrafts or be promoted toward the surface by strong upflows.

It is also clear that the buoyant flux tubes discussed in this section attempt to achieve a state of horizontal equilibrium early in their evolution.  Especially in the upper convection zone, the initial imbalance of horizontal forces can bring the tube into shallower layers before a horizontal equilibrium is found.  In stars with strong differential rotation, assuming the tubes are built co-rotating with the local plasma, this will bring the tube outward into layers with increasingly prograde motion.  While the azimuthal drag force has no effect on axisymmetric flux tubes in our formulation (other than the prescribed $v_{\phi 0}$ here), tubes that develop distinct buoyantly rising loops due to convective motions could be pushed prograde through the drag force acting on the loop legs.  This additional supply of angular momentum will reduce the poleward deflection of the rising loop, and may help to achieve lower latitude flux emergence \citep[see e.g.][]{fan1994}.  In Section \ref{sec:tubes_conv}, we will assume that the flux tube is built co-rotating at the same rate as the surrounding plasma, adopting both a \emph{fast} differential rotation profile as well as a \emph{slow} profile rotating closer to the solid body rate (see Fig. \ref{fig:diffrotprof}b).  The latter is predicted by 3D MHD simulations of dynamo action in fully convective stars and inferred from observations of low-mass stars.








\subsection{Effects of Radiative Heating}
\label{sec:radheat}

In reality, it is likely that flux tubes neither rise perfectly adiabatically nor adjust instantly to the temperature of their surroundings.  Rather, there is a rate at which heat flows in or out of the tube, given here by Eq. \ref{eqn:rad}.  \citet{weber2015} have shown that additional heating of the flux tube in the lower convection zone provided by the deviation in the mean temperature gradient from radiative equilibrium (i.e. $\nabla \ne \nabla_{rad}$) can significantly enhance the buoyancy of flux tubes.  Additionally, there is a diffusion of heat across the flux tube due to the temperature differences between the flux tube and the external plasma.  Owing to the lower thermal diffusivity ($\kappa_{diff}=\kappa_{e}/\rho c_{p}$) in fully convective stars compared to earlier spectral types, the radiative diffusion timescale $\tau_{2}=a^{2}/\kappa_{diff}$ (corresponding to the second term on the right hand side of Eq. \ref{eqn:rad}) across a tube of radius $a\sim10^{8}$ cm is of order $\sim10^{12}$ s at $0.5$R, much longer than the rise times of the flux tubes in our simulation by $\sim4-5$ orders of magnitude (see Section \ref{sec:diff}), and is therefore negligible.  Since we do include a radiative heating term in our model, we briefly examine its effects here.  

We can assess the relative importance of radiative heating by comparing the growth of the buoyancy $\Delta \rho = \rho_{e}-\rho$ caused by radiative heating to that from the adiabatic expansion of the flux tube rising through a superadiabatically stratified medium \citep[see][]{fan1996}:
\begin{eqnarray}
\bigg( \frac{d \Delta \rho}{dt} \bigg)_{rad} & = & \frac{\rho_{e}}{p_{e}}\nabla_{ad} (-\nabla \cdot {\bf F}_{rad}), 
\label{eq:drho_rad} \\
\bigg( \frac{d \Delta \rho}{dt} \bigg)_{ad} & = & \rho_{e} \frac{v_{r}}{H_{p}} \delta, 
\label{eq:drho_ad}
\end{eqnarray}
where $\delta=\nabla-\nabla_{ad}$, the excess of the background temperature gradient above the adiabatic value, and $v_{r}$ is the radial velocity of the flux tube cross-section.  The contribution to the buoyancy evolution from radiative heating is dependent on radius only, and therefore the same for all flux tubes at the same depth.    

\begin{figure}
\centering
\includegraphics[scale=.45,clip=true,trim=.5cm 0cm 0cm 0cm]{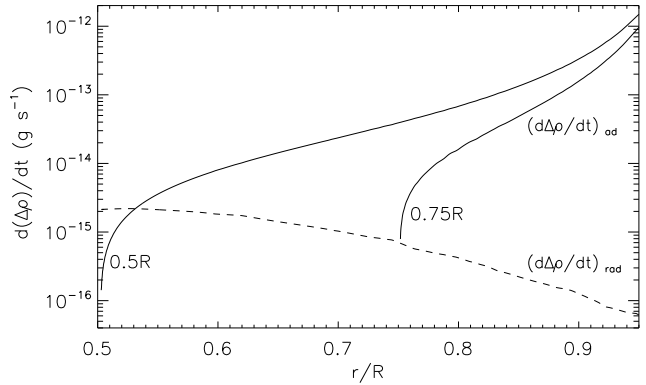}
\caption{Growth of buoyancy resulting from adiabatic expansion of Case THE, $B_{0}=30$ kG, $\theta_{0}=5^{\circ}$ flux tube cross-sections (Eq. \ref{eq:drho_ad}, solid lines) compared to that resulting from radiative heating (Eq. \ref{eq:drho_rad}, dashed line).  Flux tube evolution is adiabatic across most of the computational domain.  Solid curves are plotted once the tube has moved radially 0.1 Mm ($\sim$0.002R) from $r_{0}$.}
\label{fig:radheat}
\end{figure}

Figure \ref{fig:radheat} depicts the contributions to the buoyancy evolution of the flux tube from both radiative heating and adiabatic expansion (Eq. \ref{eq:drho_rad} and \ref{eq:drho_ad}, respectively).  As flux tubes with weaker magnetic field strengths are less buoyant (i.e. smaller $\Delta \rho$), the boost to the buoyancy evolution from the uniform radiative heating has a comparatively stronger effect.  In the upper 75$\%$ of the convection zone, radiative heating has minimal influence on the buoyancy evolution of the flux tube.  For very low latitude $\theta_{0}=1^{\circ}$, 30 kG flux tubes, radiative heating reduces the $\sim$325 day rise time of the adiabatically evolving case (Case ATLf) by $\sim$10 days (Case TLf).  The rise of these same flux tubes initiated at 0.5R are reduced by at most $\sim$140 days (Case TLf) from the $\sim$700 day (Case ATLf) adiabatic rise time.  The majority of the buoyancy increase from radiative heating for tubes at this depth occurs across a short distance of $\sim$0.03R upward from 0.5R (see Fig. \ref{fig:radheat}).  At both depths, radiative heating has a negligible effect on the rise of $100-200$ kG flux tubes.  While the incorporation of radiative heating to the model does change the rise time compared to the adiabatically evolving case in some circumstances, we find that the oscillations and horizontal force balance discussed in Section \ref{sec:dynamic} are largely unaffected.  

As a technical detail, in order to plot Equation \ref{eq:drho_ad} at the flux tube cross-section on a y-log axis, it is necessary to eliminate the radial oscillations the flux tube executes.  We achieve this by performing simulations where the initial flux tube is in horizontal force balance between the Coriolis, buoyancy, and tension forces.  This entails adopting a slightly modified initial azimuthal velocity
\begin{equation}
v_{\phi he} = \frac{\frac{B^{2}}{4 \pi r_{0} \sin{\theta_{0}}}-\frac{B^{2}}{8 \pi} \frac{\sin{\theta_{0}}}{H_{p}}}
{2 \Omega_{0} \rho_{e} (1-\frac{B^{2}}{8 \pi p_{e}})}.
\label{eq:vphi}
\end{equation} 
By eliminating the oscillations, the rise time and trajectory of the flux tube are changed somewhat, however the relevant conclusion remains the same: namely that the evolution of even the weakest flux tube we consider is mostly adiabatic, with the exception of a short distance across the deep interior.  

For this purposes of this paper, we are mostly interested in assessing how turbulent convective flows influence the buoyant rise of active-region-scale flux tubes in a fully convective star.  In the following sections, we will include radiative heating in all simulations, comparing flux tubes rising in a quiescent convection zone to those allowed to evolve under the influence of convective flows.                           

\subsection{A Note on Flux Tubes Initially in Mechanical Equilibrium}
\label{sec:mecheq}

As mentioned in Section \ref{sec:init}, flux tubes originating in an isothermal radiative interior will cool as they rise through the region, achieving neutral buoyancy.  If the tube is located outside the equatorial plane, the poleward component of the unbalanced magnetic tension will force the tube to move latitudinally, closer to the rotation axis.  This in turn induces a prograde flow of plasma inside the tube due to the conservation of angular momentum.  Eventually a state will be reached where the buoyancy force vanishes and the inward directed magnetic tension is balanced by the now outward directed Coriolis force.  This is the state of mechanical equilibrium  (hereafter MEQ), wherein all forces acting on the neutrally buoyant flux tube in all directions have come to a state of equilibrium \citep[see e.g.][]{moreno1992}.  The flux tube may then execute oscillations around this equilibrium, with a large body of work devoted to studying the stability of flux tubes in MEQ under a variety of conditions and geometries \citep[e.g.][]{spruit1982,ferriz1993,ferriz1995,caligari1995}.    

\begin{figure}
\centering
\includegraphics[scale=.52,clip=true,trim=1cm 0cm 1cm 0cm]{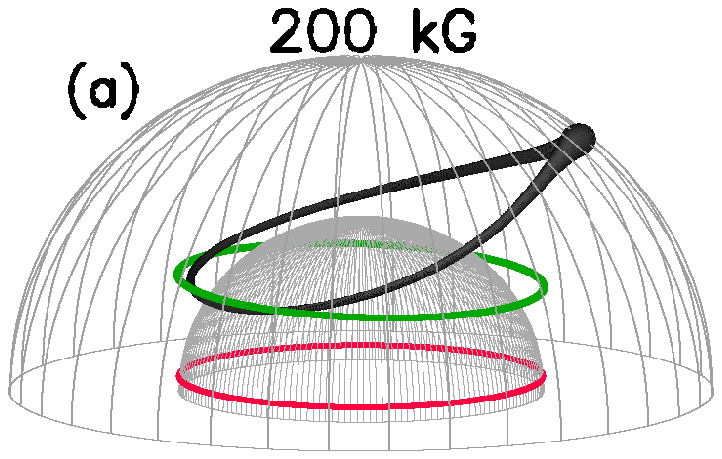}
\includegraphics[scale=.52,clip=true,trim=1cm 0cm 1cm 0cm]{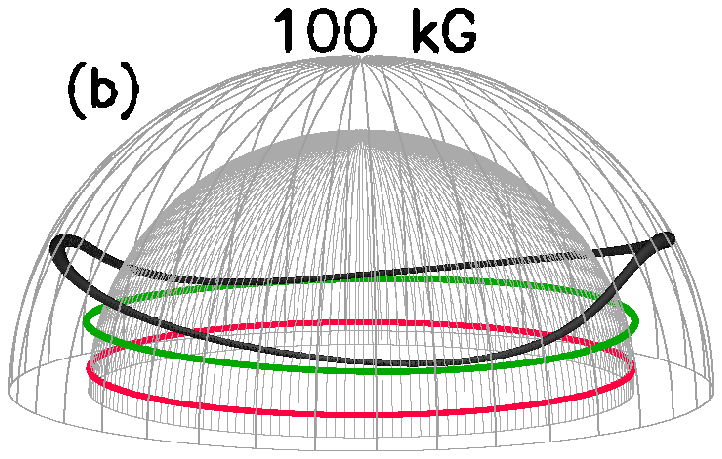} 



\caption{Time evolution of flux tubes initially in mechanical equilibrium (Case M) with $\theta_{0}=5^{\circ}$.  Plotting specifics are the same as Figure \ref{fig:no_conv}, except the azimuthal axis has been rotated so the apex of the tube at the upper boundary is on the right-hand side.  Flux tubes originally in mechanical equilibrium can develop non-axisymmetric undular instabilities ($m\ne0$) if the magnetic field strength and radius of curvature is large enough.}
\label{fig:no_conv_mecheq}
\end{figure}

\begin{figure*}
\centering
\includegraphics[scale=.6,clip=true,trim=0cm 0cm 1cm 0cm]{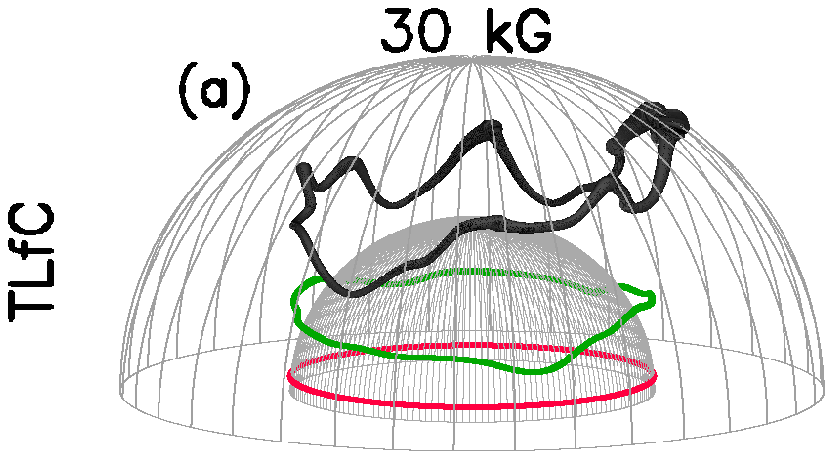}
\includegraphics[scale=.6,clip=true,trim=0cm 0cm 1cm 0cm]{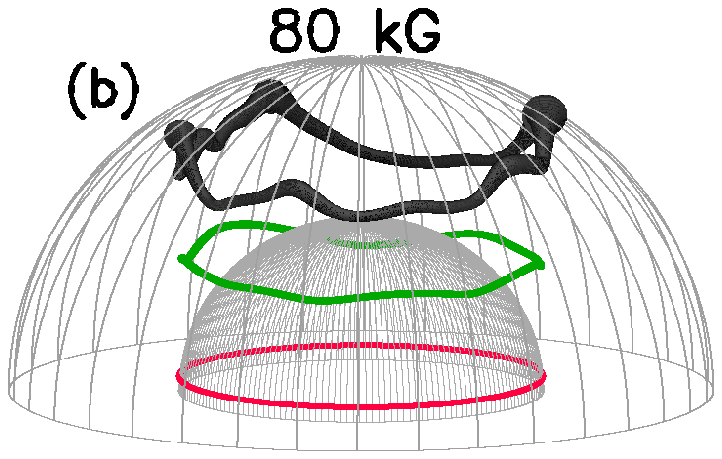}
\includegraphics[scale=.6,clip=true,trim=0cm 0cm 1cm 0cm]{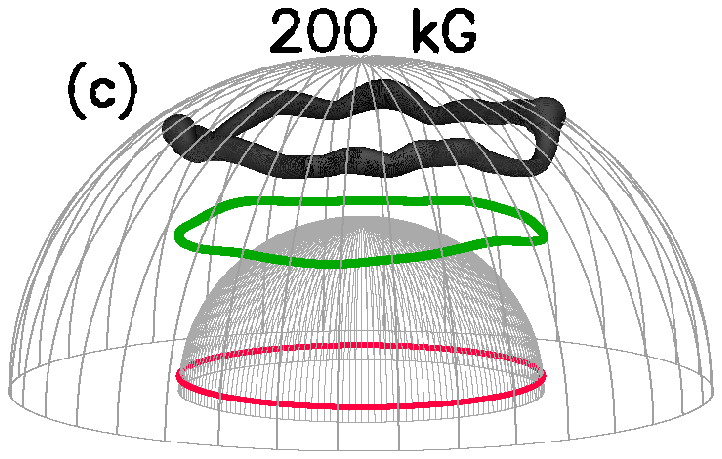} \\
\includegraphics[scale=.6,clip=true,trim=0cm 0cm 1cm 0cm]{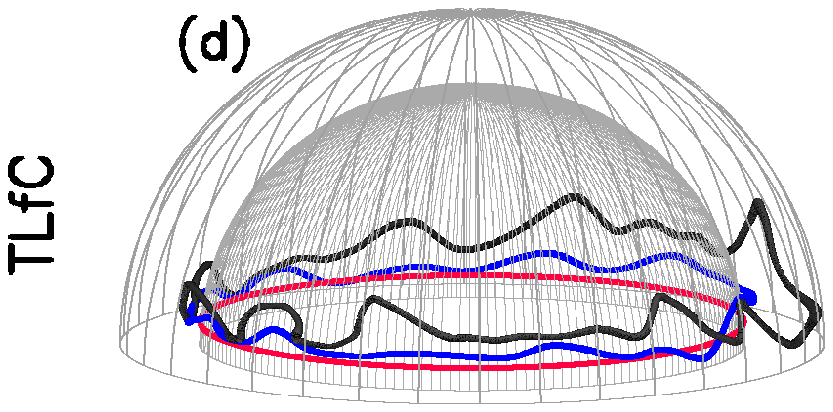}
\includegraphics[scale=.6,clip=true,trim=0cm 0cm 1cm 0cm]{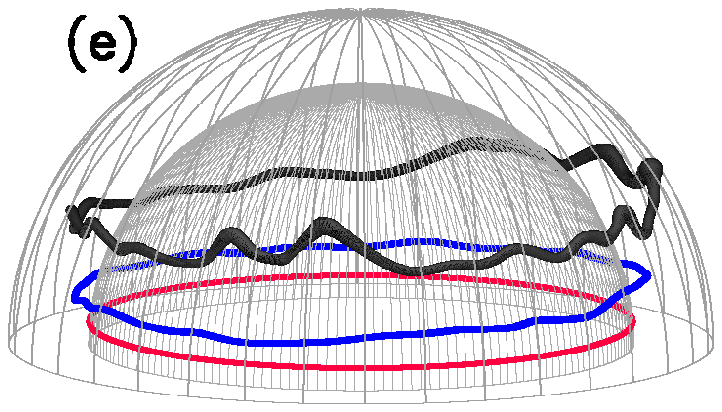}
\includegraphics[scale=.6,clip=true,trim=0cm 0cm 1cm 0cm]{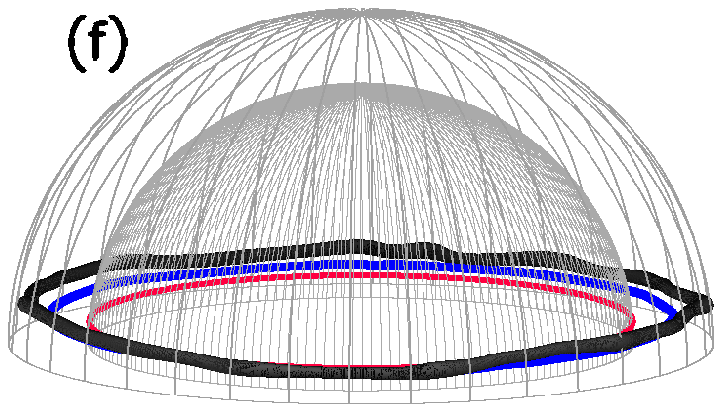} \\
\includegraphics[scale=.6,clip=true,trim=0cm 0cm 1cm 0cm]{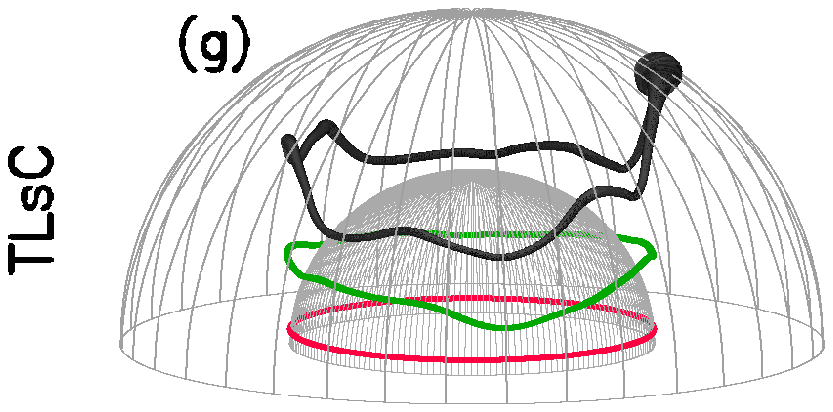}
\includegraphics[scale=.6,clip=true,trim=0cm 0cm 1cm 0cm]{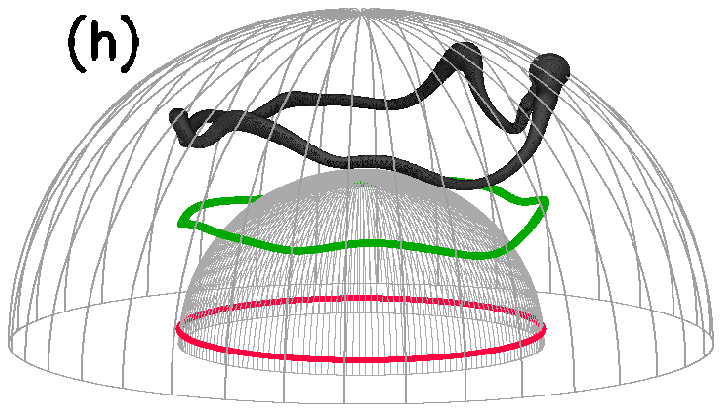}
\includegraphics[scale=.6,clip=true,trim=0cm 0cm 1cm 0cm]{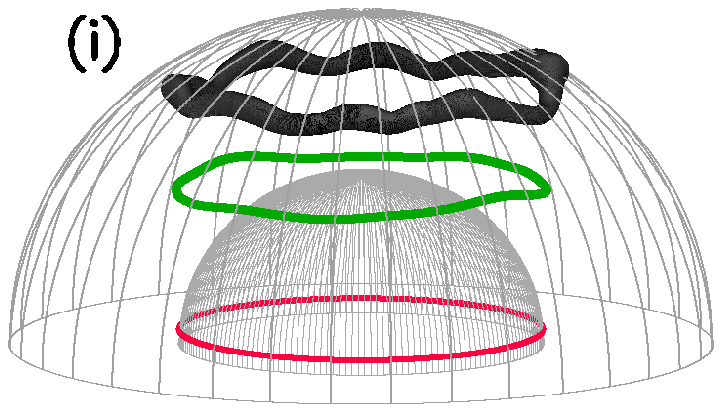} \\
\includegraphics[scale=.6,clip=true,trim=0cm 0cm 1cm 0cm]{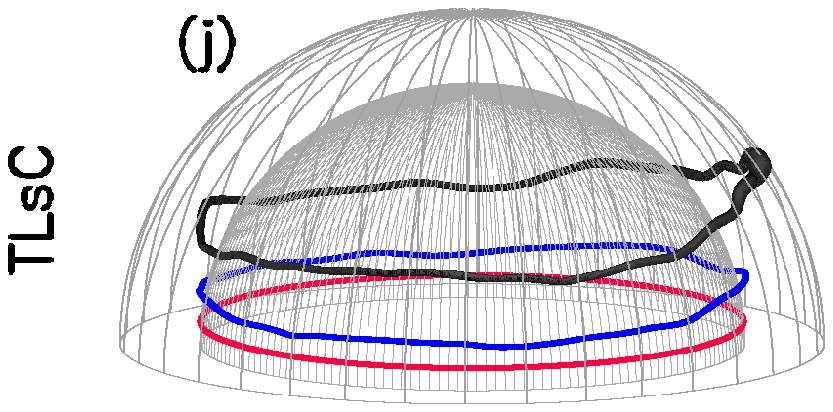}
\includegraphics[scale=.6,clip=true,trim=0cm 0cm 1cm 0cm]{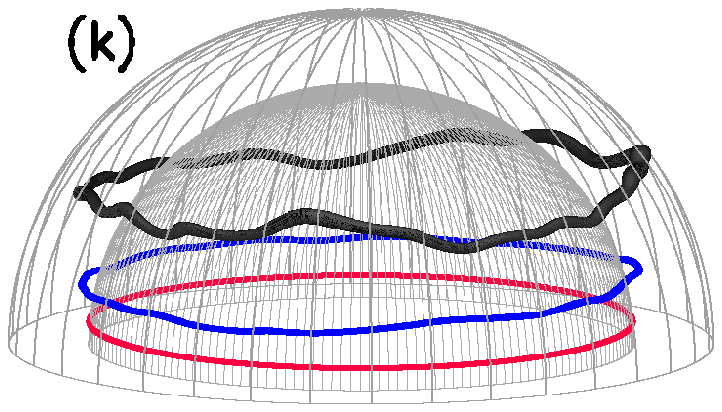}
\includegraphics[scale=.6,clip=true,trim=0cm 0cm 1cm 0cm]{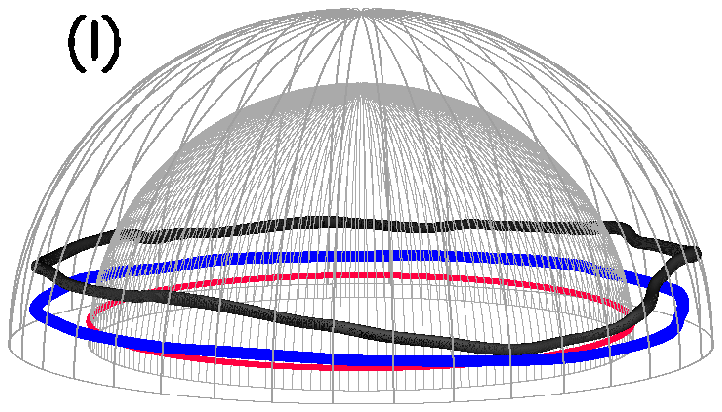} \\






\caption{Case TLfC ({\bf a}-{\bf f}) and TLsC ({\bf g}-{\bf l}) flux tubes initiated at $\theta_{0}=5^{\circ}$ with initial depths of 0.5R and 0.75R, represented by the inner mesh sphere of constant $r_{0}$.  The magnetic field strength $B_{0}$ of the flux tube is the same for each column.  The tube is plotted at three different instances, with the colors corresponding to those given in Figure \ref{fig:no_conv}. As in Figure \ref{fig:no_conv_mecheq}, the image has been rotated so the apex of the tube at the upper boundary is on the right-hand side.  Convection modulates the rise of the flux tube, promoting buoyantly rising loops.}
\label{fig:convection_tubes}
\end{figure*}

In the Solar context, perturbations to flux tubes in MEQ can result in undular magnetic buoyancy instabilities (i.e. $m=1-3$).  Portions of the tube then anchor in the overshoot region as material drains from the crests to the troughs, promoting buoyantly rising loops and sinking of the troughs into the convectively stable interior \citep[e.g.][]{caligari1998,weber2011}.  However, if the radius of curvature is too small, as would be the case if the tube is initiated at high latitudes or near the radiative zone/convection zone interface in stars with small radiative cores, the tube may still slip poleward if $m=0$ (axisymmetric flux ring) is the fastest growing unstable mode, driven by a dominant magnetic tension force compared to buoyancy \citep[e.g.][]{granzer2000,holzwarth2001}.  We find a similar behavior for $B_{0}\le80$ kG tubes initiated at 0.75R, and $B_{0}\le100$ kG initiated at 0.5R, if the initial condition of MEQ (Case M) is applied.  In these simulations, we have included radiative heating following the formulation given in Section \ref{sec:TFTeq} and used the cross-sectional radius $a_{0}=$1.7\e{8} cm.  Above this threshold, the flux tubes develop significant undular instabilities with low order unstable modes if the radius of curvature is large enough.  Figure \ref{fig:no_conv_mecheq} shows the time evolution of two Case M flux tubes rising through the 0.3M$_{\odot}$ convective envelope that have developed dominant $m=1$ (Fig. \ref{fig:no_conv_mecheq}a) and $m=2$ (Fig. \ref{fig:no_conv_mecheq}b) undular instabilities.     


We think it more likely that flux tubes are built in fully convective stars in a state of TEQ rather than MEQ, as discussed in Section \ref{sec:init}.  The choice between these two initial conditions will alter the subsequent evolution of the flux tubes and balance between the relevant forces, having significant effects on properties such as rise times, latitude of emergence, and development of $\Omega$-shaped loops, or lack thereof.  Depending on the initial latitude, depth, or magnetic field strength, flux tubes initiated in MEQ can take at most an order of magnitude longer to rise than their counterparts initiated in TEQ.  While tubes in TEQ are immediately buoyant with a density deficit following Equation \ref{eq:tempeq}, those in MEQ are subject to the growth rate of the magnetic buoyancy instability.  The apices then rise toward the surface as bending of the tube drains material out of the apex to the trough, further depleting the density there and subsequently increasing the buoyancy. 

The choice of TEQ versus MEQ may also change the rise times reported in \citet{browning2016} somewhat for the extreme magnetic fields of $10^{6}-10^{7}$ G.  In that paper, we performed the TFT simulations primarily to confirm that the rise time varies inversely to the cross-sectional radius of the flux tube (see Fig. 5 in that paper).  This result is robust, and independent of the choice of MEQ or TEQ at moderate rotation rates (i.e. $\le10\Omega_{0}$) for the magnetic fields of $10^{6}-10^{7}$ G studied. 

For the parameter space explored here, we note that flux tubes initialized in both TEQ and MEQ exhibit strong poleward deflection, even at the very modest Solar rotation rate.  As pointed out in Section \ref{sec:dynamic}, our flux tubes in TEQ do exhibit some degree of radial motion (see Figure \ref{fig:choud_plot}) ultimately determined by the horizontal forces.  Once a balance in this direction is achieved, the flux tubes then turn poleward.  This is in stark contrast to the radial trajectories of flux tubes initialized in MEQ in the quiescent Solar convection zone described in \citet{weber2011,webersolphys2013}, especially for the strongest magnetic field strengths of 60-100 kG.  In those papers, the majority of flux tubes develop undular instabilities, resulting in troughs that effectively `anchor' in the sub-adiabatic overshoot region.  This keeps the flux tube from migrating too far poleward before a buoyantly rising loop reaches the near-surface region.  If $m=0$ is the fastest growing unstable mode, the tube may not anchor, freely floating with motion parallel to the rotation axis and emerging at higher latitudes than expected, as shown in \citet{weber2011} for some weaker $B_{0}$ flux tubes. 

Futhermore, as previously mentioned, the choice of TEQ or MEQ will make a difference in the dynamic evolution of forces acting on the flux tube.  The TFT simulations of \citet{fan1993,fan1994}, for instance, begin with flux tubes in TEQ in a Solar convection zone, but must embed portions of them in a strongly sub-adiabatic overshoot region to reinforce anchoring.  Comparing the TEQ flux tubes of \citet{fan1993} and MEQ flux tubes of \citet{webersolphys2013}, the emergence latitudes of buoyantly rising loops are fairly commensurate, but can be larger by up to $10^{\circ}$ for the B$_{0}=30-50$ kG, $\theta_{0}\le5^{\circ}$ TEQ flux tubes of \citet{fan1993}.  In the Solar context, \citet{caligari1998} discuss in detail how the choice between TEQ and MEQ effects the anchoring of the flux tube, geometrical asymmetries of the rising loop, and the asymmetry between the magnetic field strength in the leading and following legs of the rising loop.



\section{Flux Tubes in Turbulent Convection}
\label{sec:tubes_conv}

\begin{figure*}

\vspace{.04 \textwidth}

\epsscale{.98}
\plotone{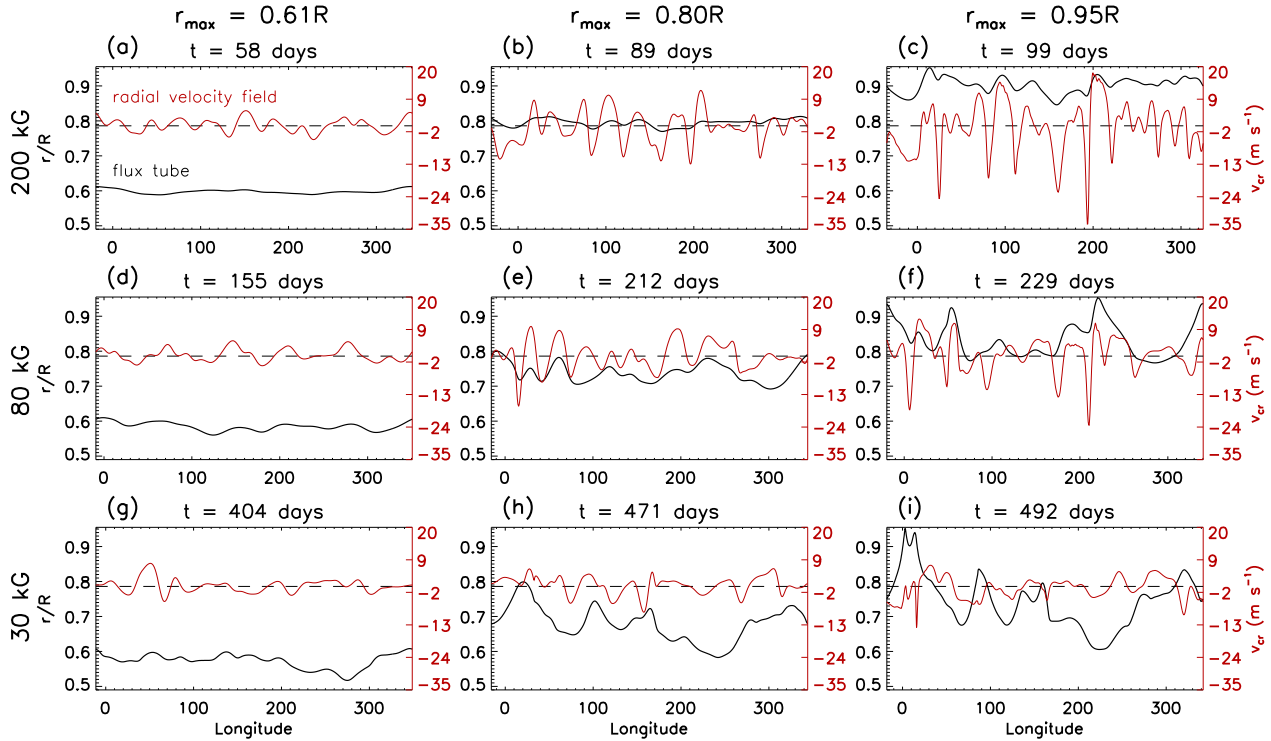} 

\caption{Evolution of Case TLfC flux tubes (black line, left axis) in the $r-\phi$ plane with initial latitude $\theta_{0}=5^{\circ}$, $r_{0}=0.5$R, and $B_{0}$ decreasing from top to bottom.  Snapshots are taken at times as indicated when the apex has reached a height of (left column) 0.61R, (middle column) 0.8R, and (right column) 0.95R, corresponding to the last timestep and the black flux tubes shown in the top row of Figure \ref{fig:convection_tubes}.  All flux tubes shown are initialized at the same time, and therefore experience the same initial flow field.  Also plotted is the external radial velocity field $v_{cr}$ (red line, right axis) at the same instant acting on each flux tube segment.  The dashed line represents zero on the $v_{cr}$ axis.  This Figure clearly shows how strong downflows can modulate the shape of the flux tube.}
\label{fig:vexrandr_0.5R}
\end{figure*}

\subsection{Qualitative Description}
The journey of an active-region-scale flux tube from its region of generation to the near-surface is in part shaped by the local and mean flows it encounters.  Previous works have sufficiently demonstrated that strong downflows can pin portions of the flux tube to deeper layers, while upflows may aid in boosting portions toward the surface \citep[e.g.][]{fan2003,jouve2009,weber2011,nelson2013}.  In the remainder of this section, we qualitatively outline how convective motions influence the flux tubes we simulate here.  

Figure \ref{fig:convection_tubes} shows 3D snapshots of representative flux tubes initiated at 0.5R and 0.75R at multiple times during their evolution.  The top two rows (Fig. \ref{fig:convection_tubes}a-\ref{fig:convection_tubes}f) exhibit flux tubes that have evolved subject to the \emph{fast} differential rotation profile (TLfC), with the bottom two rows (Fig. \ref{fig:convection_tubes}g-\ref{fig:convection_tubes}l) subject to the \emph{slow} differential rotation profile (TLsC), where the tubes are assumed to have been built co-rotating with the surrounding plasma. All flux tubes in Figure \ref{fig:convection_tubes} are initialized at the same time, and therefore experience initially the same time-varying radial and latitudinal flows. Figure \ref{fig:vexrandr_0.5R} complements the top row of Figure \ref{fig:convection_tubes}, depicting the time evolution of the same flux tubes in the $r-\phi$ plane, as well as the radial flows acting on each flux tube segment. We have chosen to present Figure \ref{fig:vexrandr_0.5R} only for representative Case TLfC tubes initiated at 0.5R as an example of how radial convective flows modulate the initial axisymmetric shape of the flux tube.  

For flux tubes of similar cross-sectional radius $a$, the severity of the distortion by convective flows increases with decreasing magnetic field strength (i.e. from right to left in Fig. \ref{fig:convection_tubes}, top to bottom in Fig. \ref{fig:vexrandr_0.5R}).  Simply, tubes of larger magnetic field have a greater magnetic tension and a stronger buoyancy force compared to the aerodynamic drag imparted by the turbulent flows, rendering the tube less susceptible to convection.  This means that tubes of larger $B_{0}$ evolve more like the axisymmetric flux tubes in a quiescent medium described in Section \ref{sec:dynamic}.      

Imprinted upon the shape of the flux tube at any moment is both the history of convective flows it has encountered as well as artifacts from more recent flows.  Small perturbations to the tube from flows in the deeper interior may continue to grow as material drains from a peak along the tube to a trough.  The peak will continue to rise toward the surface, sometimes being boosted by an upflow, while other times pummeled by a strong downflow.  For example, the 30 kG Case TLfC flux tube initiated at 0.5R, depicted in Figure \ref{fig:convection_tubes}a and the bottom row of Figure \ref{fig:vexrandr_0.5R}, develops small undulations in the deep interior that grow as the peaks along the tube rise into less dense layers.  By the time the fastest rising peak reaches the upper boundary, which we will subsequently refer to as the flux tube \emph{apex}, the tube has developed multiple loops with troughs that extend to as deep as 0.6R.  A strong downflow passes across the apex of the flux tube as it nears the surface, creating a double peaked feature in the rising loop between $\sim$$0^{\circ}-20^{\circ}$ longitude in the last timestep, as shown in Figure \ref{fig:vexrandr_0.5R}i. 


Even when subjected to convective flows, both Case TLfC and TLsC flux tube initiated at a depth of 0.5R still have a mean motion that is largely parallel to the rotation axis (Fig. \ref{fig:convection_tubes}, first and third rows).  The broader, weaker radial flows in the mid-convection zone are enough to perturb flux tubes of weaker $B_{0}$, creating rising loops with troughs residing in much deeper layers.  However, at this depth, radial flows are not strong enough to bring the apex radially outward, nor is the differential rotation profile strong enough to force the loop to deviate much from parallel motion toward the poles.  As the magnetic field strength $B_{0}$ of the tube increases, the development of loop-like features only begins to occur as the ring-like tube reaches shallower regions of the convection zone where the radial motions increase in magnitude and have smaller spatial scales (see Figure \ref{fig:vexrandr_0.5R}).  Differential rotation  (and the assumed $v_{\phi 0}$ inside the tube) plays only a small role here: The evolution of the 200 kG Case TLfC and TLsC flux tubes initiated at 0.5R in Figure \ref{fig:convection_tubes}c and \ref{fig:convection_tubes}i are nearly indistinguishable.

Flux tubes originating at 0.75R can evolve differently; in particular, not all exhibit a mean motion parallel to the rotation axis.  For example, the 30 kG Case TLfC flux tube in Figure \ref{fig:convection_tubes}d and the 200 kG Case TLfC tube in Figure \ref{fig:convection_tubes}f exhibit lower latitude emergence than their counterparts evolving in the slow differential rotation profile (see Figs. \ref{fig:convection_tubes}j and \ref{fig:convection_tubes}l).  This change in evolution is solely due to the applied differential rotation profile, and likewise the initial $v_{\phi 0}$ inside the tube.  


In Section \ref{sec:diff}, we discuss in greater detail the effect the differential rotation profile has on the rise duration and emergence latitudes of buoyantly rising loops.  We investigate the ability of convection to suppress the motion of the entire flux tube in Section \ref{sec:pump}.


\subsection{Rise Times, Emergence Latitudes, and Influence of Differential Rotation}
\label{sec:diff}

\begin{figure*}
\vspace{.03 \textwidth}
\centering

\includegraphics[scale=.44]{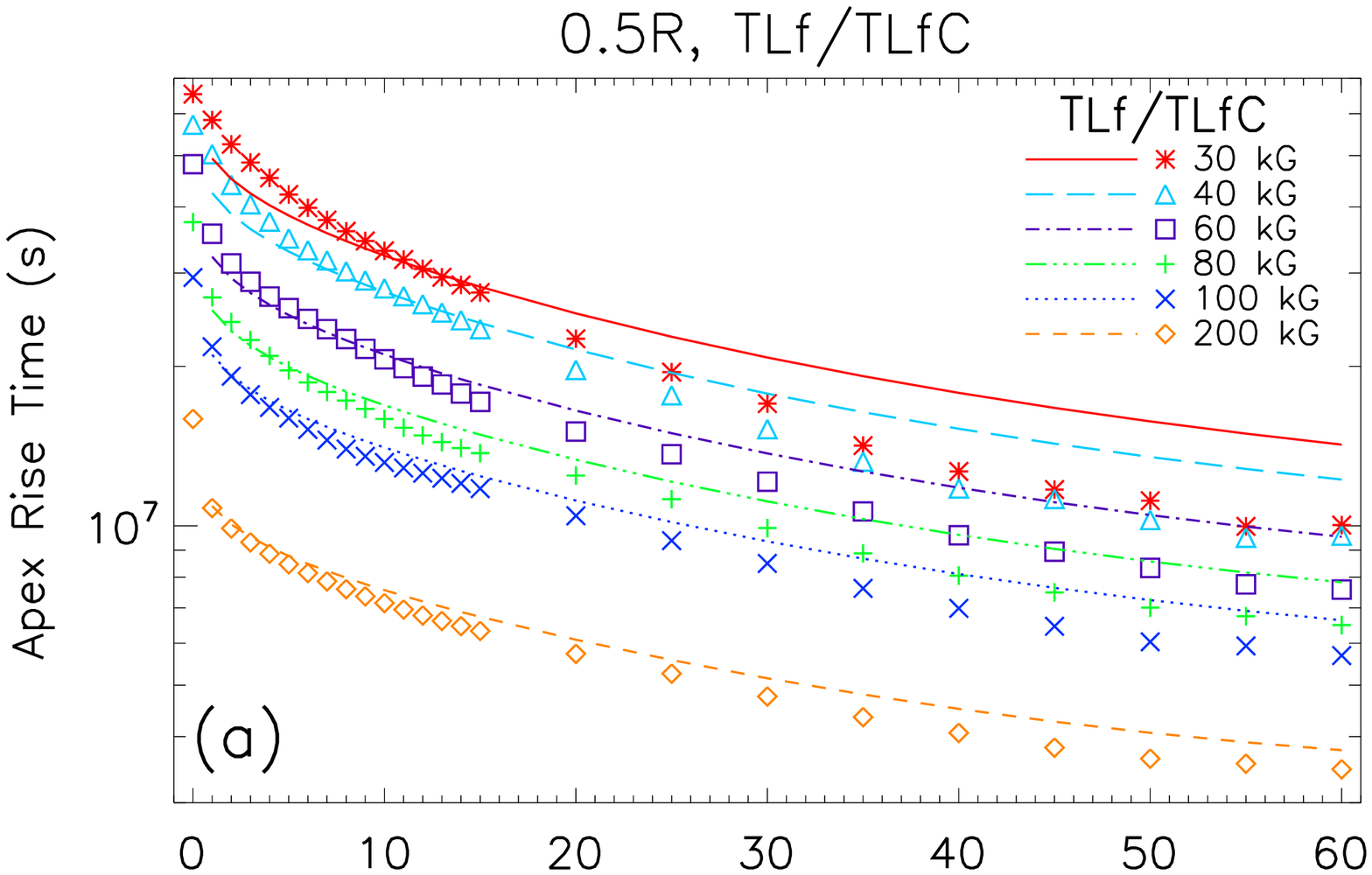}
\includegraphics[scale=.44]{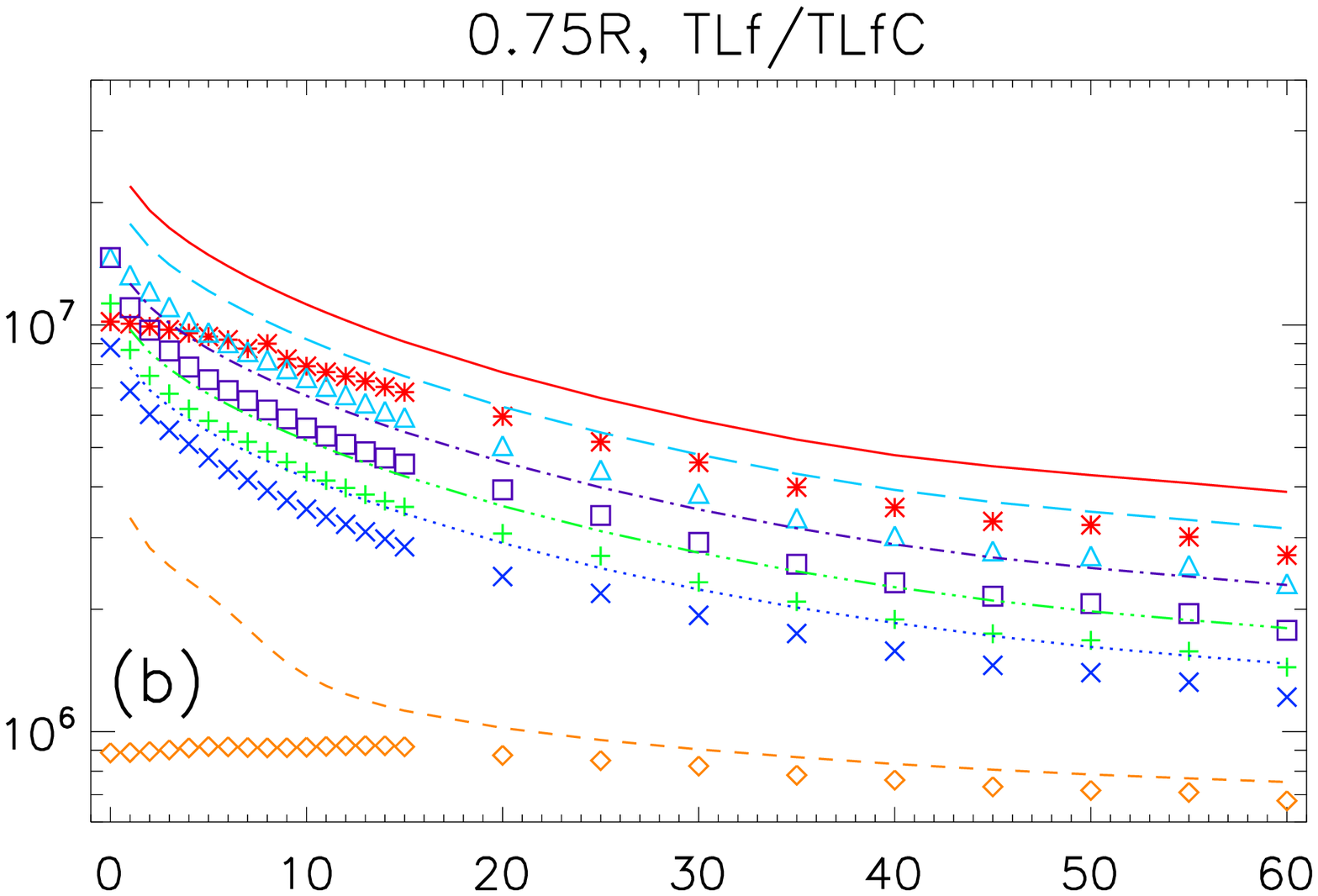}\\
\includegraphics[scale=.44]{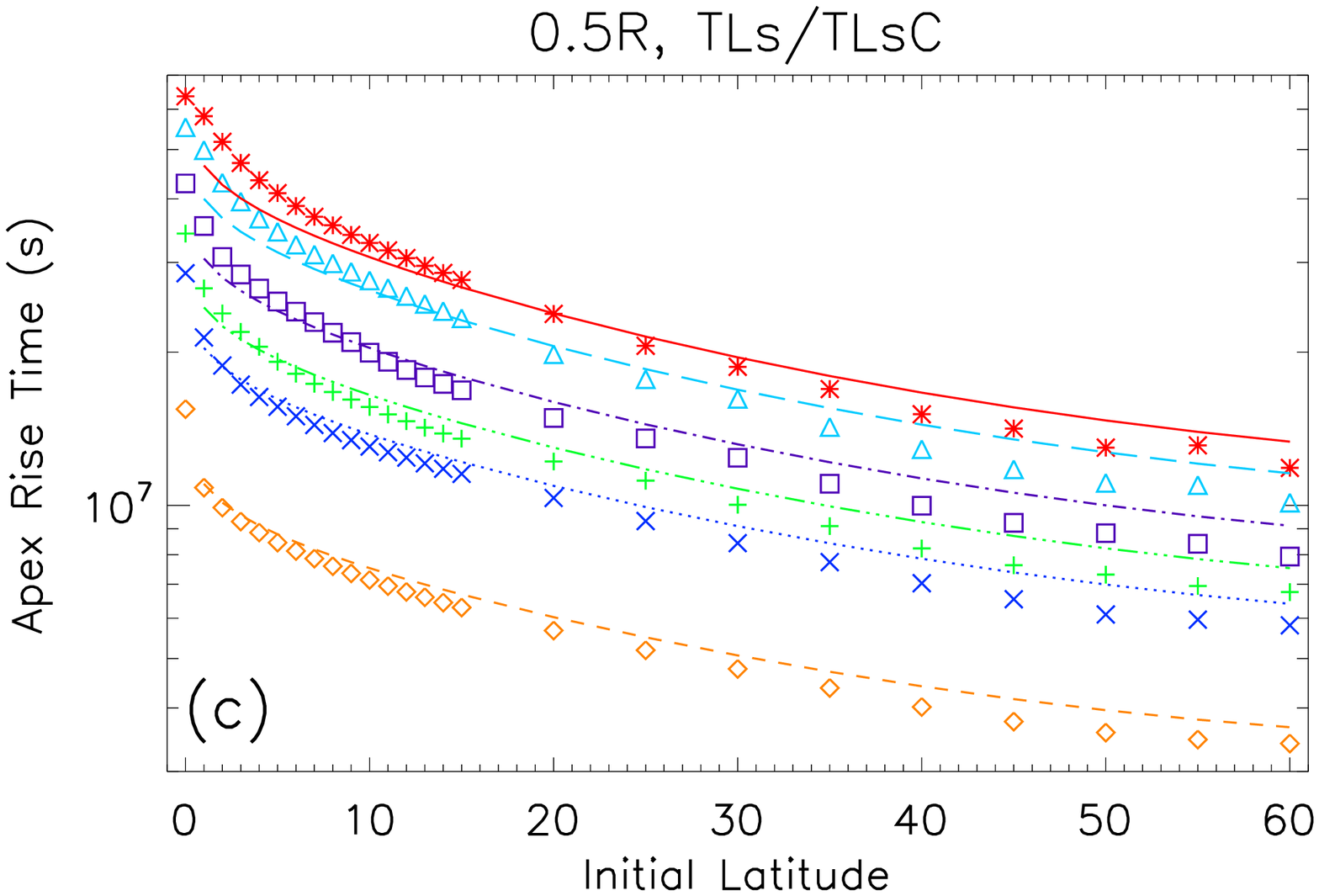}
\includegraphics[scale=.44]{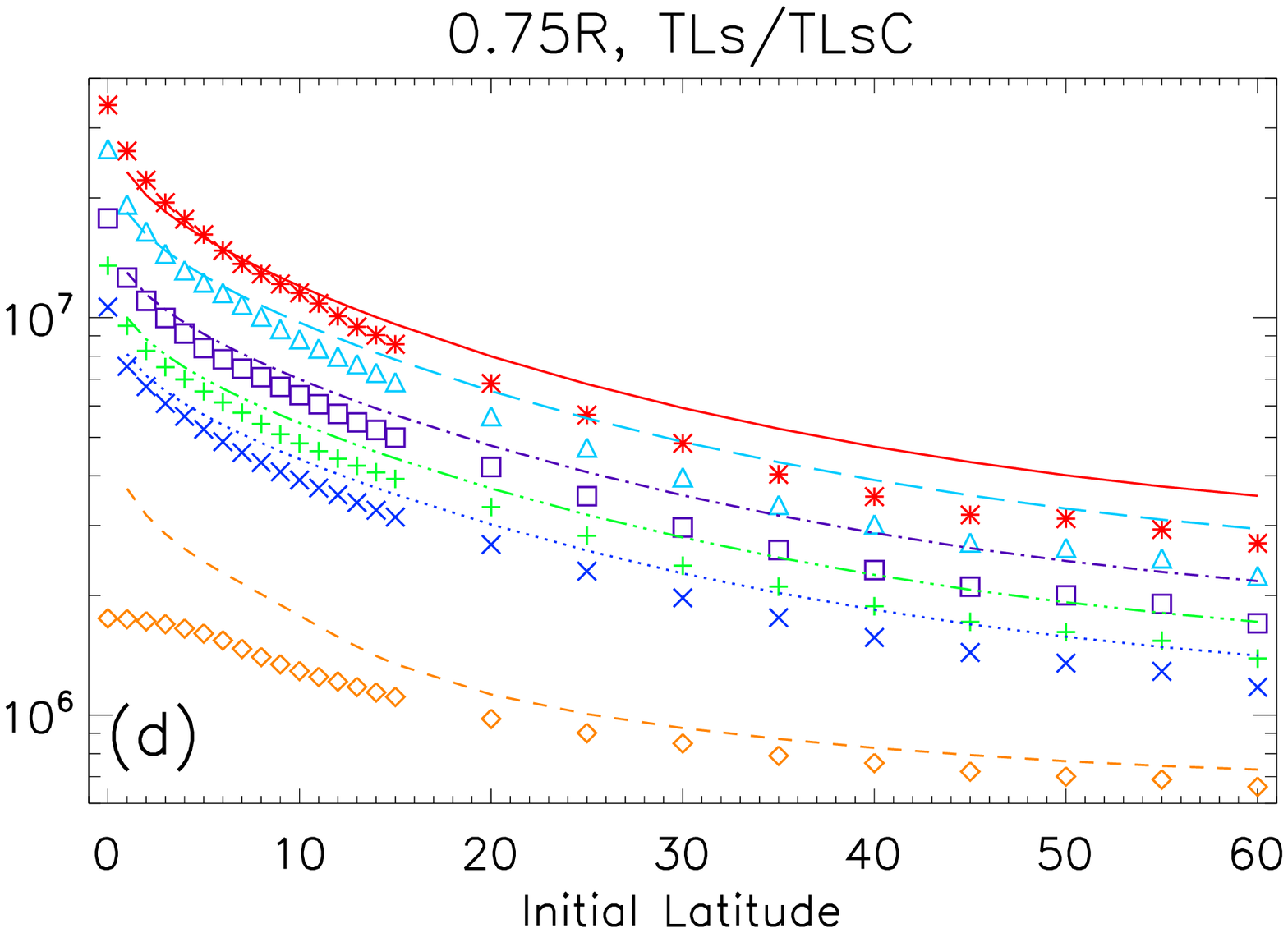}\\

\caption{Average apex rise duration for flux tubes traveling through both the quiescent interior (represented by lines) and the convective flow field (represented by symbols) as a function of the absolute value of the initial latitude.  Rise times are averaged over three unique flow fields and both hemispheres.  The standard deviation, or spread about the mean, of the average rise times is at most 15$\%$ of the average value.  A fast differential rotation profile near the equator in shallower depths aids in significantly shortening the rise time of weak $B_{0}=30-40$ kG flux tubes in the near-equatorial region.}
\label{fig:risetimes}
\end{figure*}

To a zeroth approximation, flux tubes built in TEQ in a low mass star rise parallel to the rotation axis rather than radially outward.  However, in some circumstances, initially low latitude flux tubes develop buoyant loops which can rise more radially, emerging in the near-equatorial region.  Furthermore, the mean and local flows encountered by the evolving flux tube can significantly alter the duration of a buoyantly rising loop's journey to the surface of the star.

\begin{figure*}

\vspace{.02\textwidth}

\centering
\includegraphics[scale=.44]{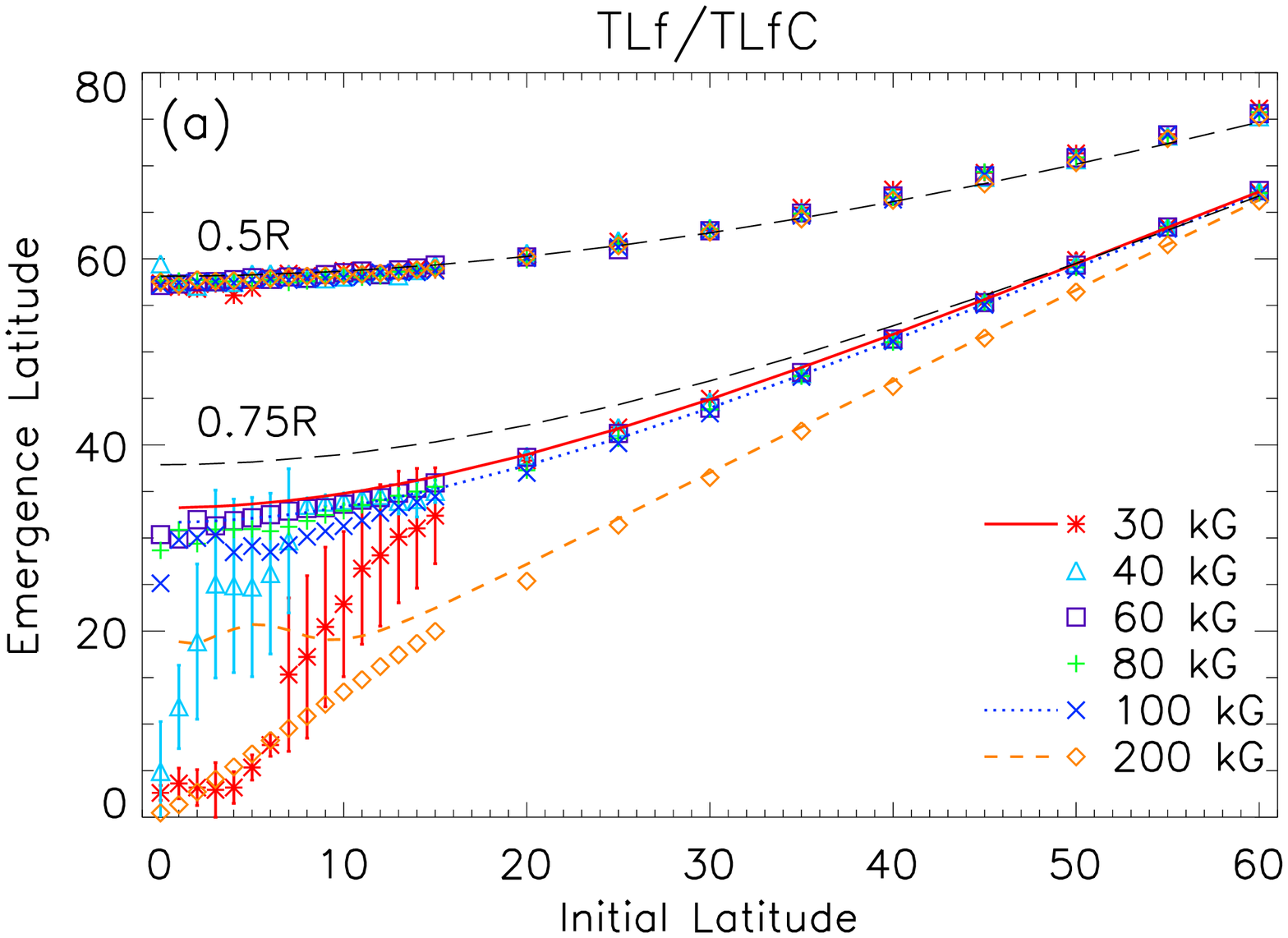}
\includegraphics[scale=.44]{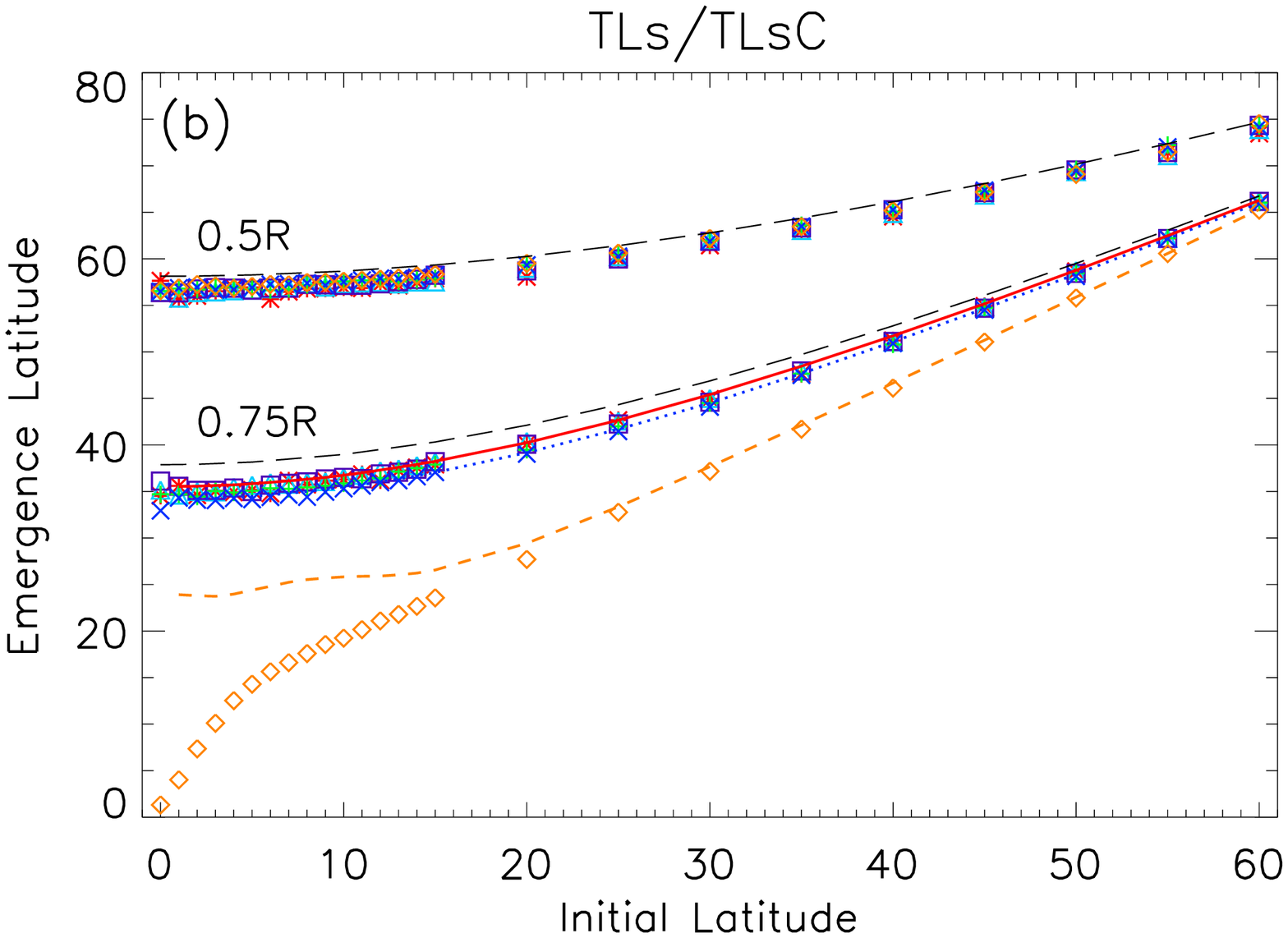}

\caption{Average emergence latitude of the flux tube apex (symbols) as a function of the absolute value of the initial latitude for Case TLfC ({\bf a}) and TLsC ({\bf b}).  The thick dashed line represents the emergence latitude if the flux tube were to rise truly parallel to the rotation axis (Eq. \ref{eqn:theta_parallel}).  Curves depicted in the legend correspond to flux tubes rising through a quiescent convection zone.  Large deviations of motion parallel to the rotation axis for the 30-40 kG Case TLfC tubes is a result of both the mean and time-varying convective flows, while the deviation for the 200 kG tubes is related strongly to the initial horizontal force imbalance and vigorous time-varying flows nearer the surface. Bars on the symbols in panel {\bf (a)} for the 30-40 kG tubes initiated at 0.75R represent the standard deviation.  The standard deviation for the other average values rarely exceeds the size of the symbols.}
\label{fig:theta_em}
\end{figure*}

Figures \ref{fig:risetimes} and \ref{fig:theta_em} show the average apex rise times and emergence latitudes, respectively, for flux tubes evolving in the time-varying $v_{r}$ and $v_{\theta}$ convective flow fields with both the fast (Case TLfC) and slow (Case TLsC) differential rotation profiles applied.  We reiterate that our flux tubes subject to convective motions are assumed to have been built co-rotating with the local differential rotation profile. Each symbol in the plots represents the quantity for all flux tubes initiated at $|\theta_{0}|$, averaged over the three ensembles we perform.  The corresponding quantities for axisymmetric flux tubes rising through a quiescent interior are also shown, and are also assumed to have been built co-rotating with the local differential rotation profile for a more consistent comparison (Cases TLf and TLs).  As our simulations terminate once some portion of the tube has reached the simulation upper boundary at 0.95R, we are only reporting the rise times and emergence latitudes for the fastest rising loops in each circumstance. 

\begin{figure}
\vspace{.02\textwidth}
\centering
\includegraphics[scale=.42]{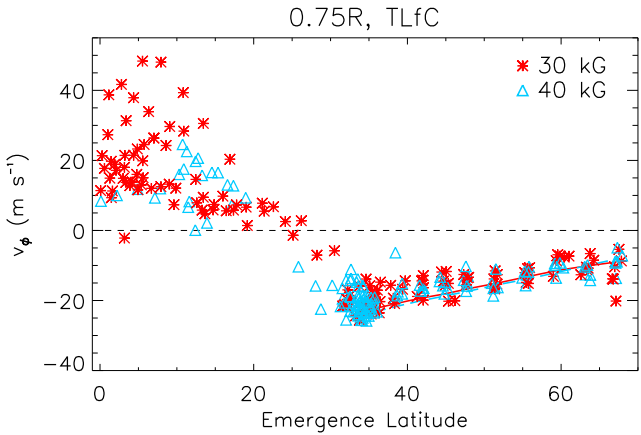}
\caption{Azimuthal velocity $v_{\phi}$ of plasma at the apex (reaching 0.95R) of $30-40$ kG Case TLfC flux tubes originating at 0.75R (symbols).  The azimuthal velocity $v_{\phi}$ along the cross-section of the axisymmetric Case TLf flux rings at 0.95R is also shown (lines).  In shallower layers, weaker flux tubes of $30-40$ kG that emerge at latitudes $\lesssim 25^{\circ}$ have developed a prograde azimuthal speed. Not shown here, all Case TLfC flux tubes of $B_{0}\ge60$ kG have an apex $v_{\phi}$ at the simulation upper boundary that is retrograde, and commensurate with conservation of angular momentum.}
\label{fig:vph}
\end{figure}

The major trend that emerges from the average rise times in Figure \ref{fig:risetimes} is the tendency in most cases for the apex rise to roughly follow the rise of the axisymmetric flux rings evolving without convection.  Weaker $B_{0}=30-40$ kG flux tubes initiated at low latitudes at 0.5R subject to both differential rotation profiles (Fig. \ref{fig:risetimes}a and \ref{fig:risetimes}c) rise slower than the same flux tubes rising through a quiescent convection zone.  This is indicative of magnetic pumping, which we further address in Section \ref{sec:pump}.  However, a stronger differential rotation profile can drastically \emph{shorten} the rise of initially low latitude $30-40$ kG (Case TLfC) flux tubes originating at 0.75R, as shown in Figure \ref{fig:risetimes}b and discussed in more detail below.   

Even though convection can modulate the tube, generating loops that extend over a large portion of the convection zone, the tendency for portions of the tube to approach horizontal equilibrium and rise parallel to the rotation axis is robust in most cases.  Figure \ref{fig:theta_em} depicts this trend.  The black dashed lines in Figure \ref{fig:theta_em} plot the emergence latitude at 0.95R expected if the flux tube rises purely vertically, following the relationship
\begin{equation}
\theta_{em}=\cos^{-1}{\bigg( \frac{r_{0} \cos{\theta_{0}}}{r_{top}} \bigg) },
\label{eqn:theta_parallel}
\end{equation}   
where $\theta$ is the latitude and $r_{top}$ is the top of the simulation domain.  For both differential rotation profiles, the 200 kG flux tubes initiated at 0.75R deviate substantially from the emergence latitude predicted by Equation \ref{eqn:theta_parallel}.  Additionally, $30-40$ kG flux tubes of $\theta_{0}\lesssim5^{\circ}-15^{\circ}$ initiated at the same depth subject to the fast differential rotation profile are able to emerge at much lower latitudes than either flux tubes evolving without convection or the prediction of Equation \ref{eqn:theta_parallel} (see panel \ref{fig:theta_em}a).       

\begin{figure*}

\vspace{.02\textwidth}

\centering
\includegraphics[scale=.46,clip=true,trim=1.cm 0cm 0cm 0cm]{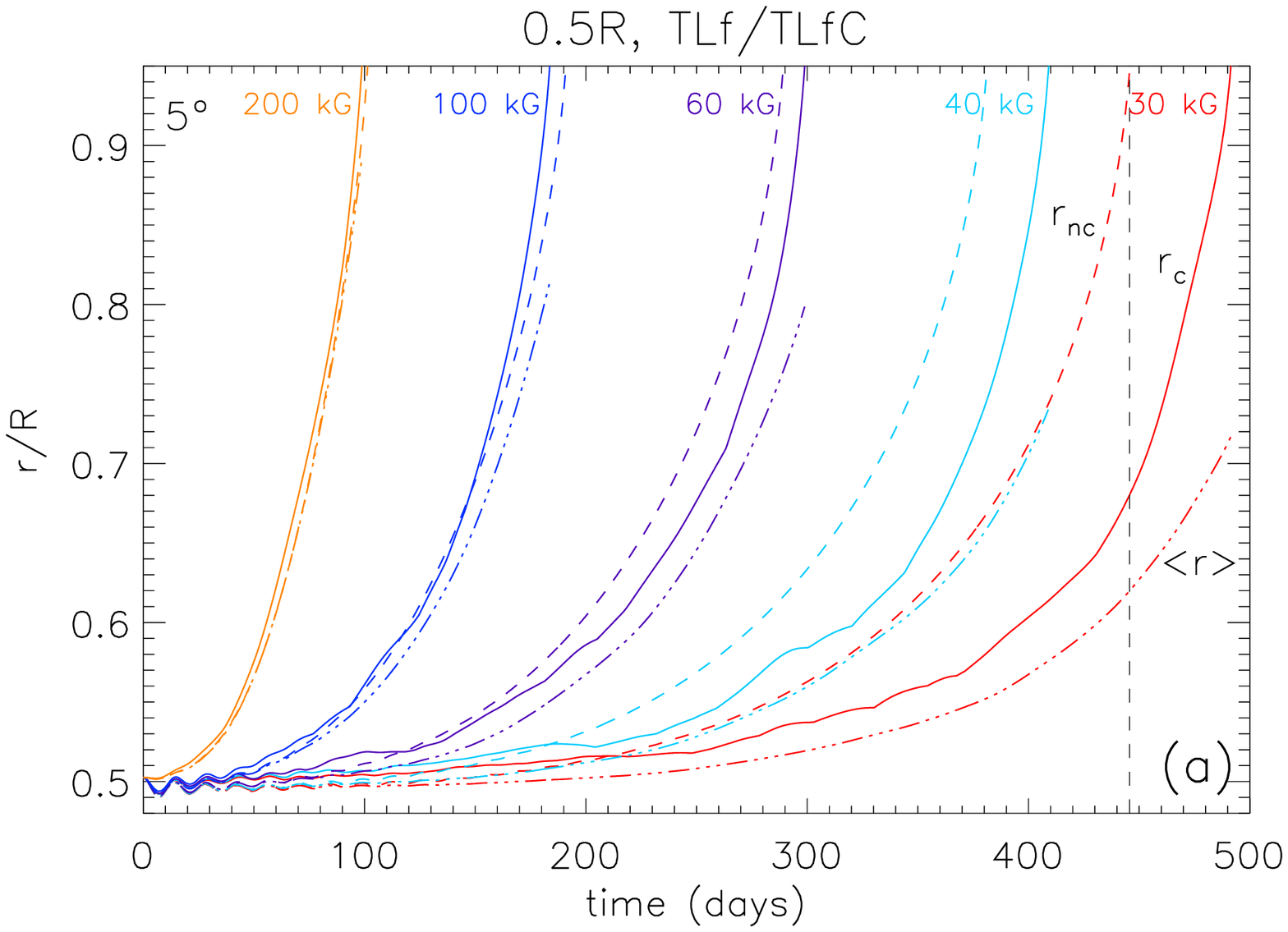}
\includegraphics[scale=.46,clip=true,trim=1.cm 0cm 0cm 0cm]{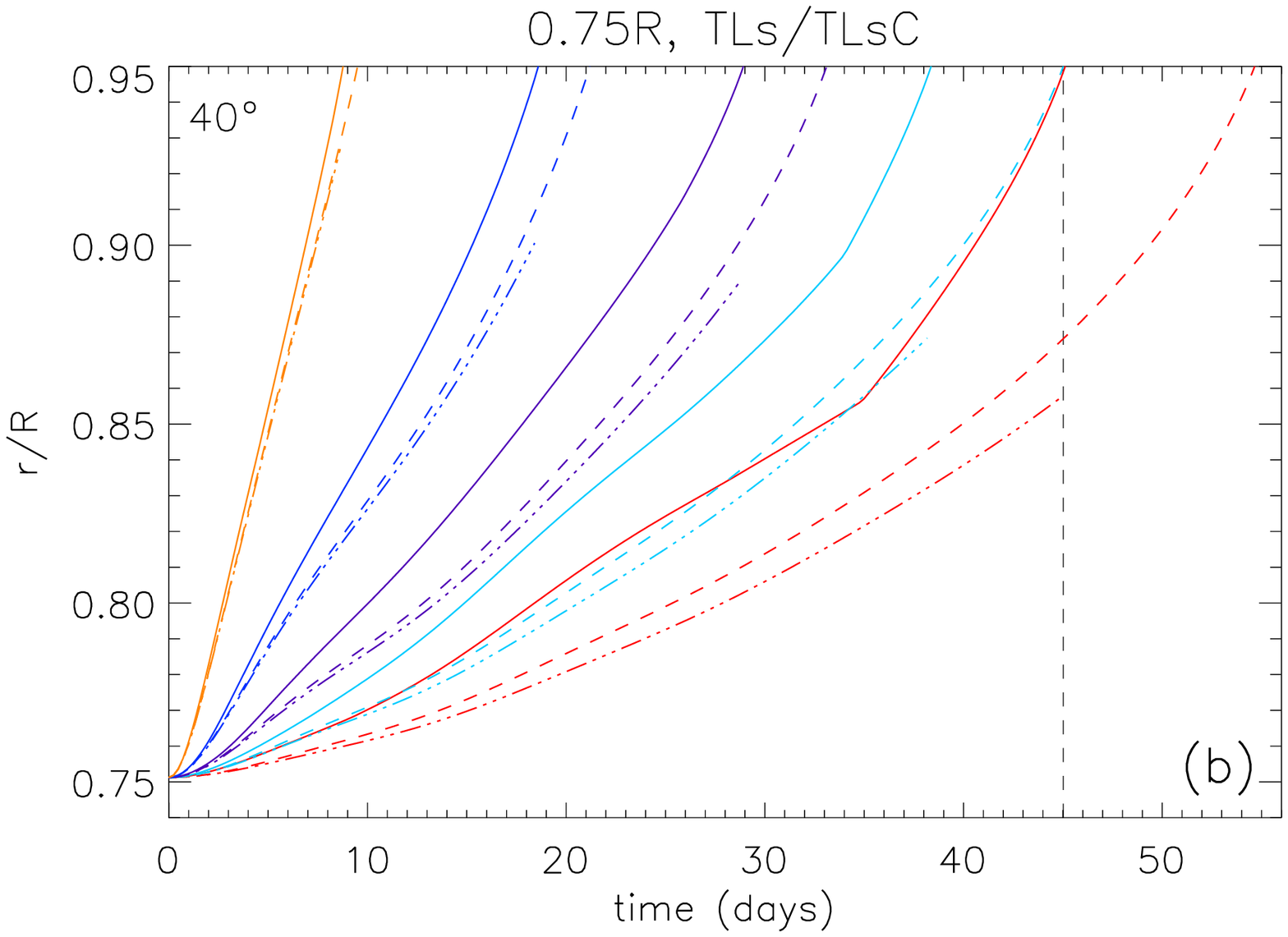}


\caption{Apex radial position $r_{c}$ (solid line) for flux tubes that evolve in convection, average magnetic field weighted radial position $\langle r \rangle$ (dash-dotted line) for the same tube, and radial position  $r_{nc}$ of the cross-section of flux tubes allowed to evolve without convection (dashed line).  These are shown for ({\bf a}) Cases TLf and TLfC tubes originating at 0.5R and $5^{\circ}$, and ({\bf b}) Cases TLs and TLsC originating at 0.75R and $40^{\circ}$.  Low latitude, weaker magnetic field strength flux tubes in the deeper convection zone are pinned down substantially by convection. Vertical dashed lines are referenced in the text (see Section \ref{sec:pump}), and correspond to $t_{min}$ for the 30 kG flux tube in the panel.}
\label{fig:pumping}
\end{figure*}

Phenomenologically, we can explain the reduced rise times and deflected latitudinal emergence of the 30-40 kG Case TLfC flux tubes by appealing to the differential rotation profile.  In the upper $\sim25\%$ of the convection zone at low latitudes, the differential rotation is strongly prograde (see Fig. \ref{fig:diffrotprof}a).  At this depth, the initial horizontal force imbalance causes the Case TLfC flux tubes to move outward into shallower layers as compared to the Case TLsC tubes owing to the larger initially outward Coriolis force (see discussion in Sec. \ref{sec:dynamic}).  Buoyant loops develop from modulation by convection.  These loops rise subject to the prograde azimuthal flow, which supplies angular momentum as the loop crosses contours of constant $\hat{v}_{\phi}$.  The retrograde flow of plasma along the flux tube expected from conservation of angular momentum will be reduced, and may even turn prograde.  A prograde azimuthal flow near the apex, if established, induces outward and equatorward components of the Coriolis force, accelerating the loop toward the surface and helping it to emerge at lower latitudes.  Furthermore, the distance the flux tube travels while executing a more radial trajectory is shorter than a trajectory parallel to the rotation axis, also reducing the rise time.  

We point out that in the absence of convection, Case TLf flux tubes initiated at 0.75R at low latitudes move horizontally outward to only $\sim$0.81R before moving parallel to the rotation axis (see Fig. \ref{fig:choud_plot}a).  The departure from the nearly parallel trajectories of 30-40 kG Case TLfC flux tubes initiated in this same region is then a result of modulation by the time-varying flows and the strongly prograde differential rotation.  To further emphasize this, Figure \ref{fig:vph} shows a scatter plot of the azimuthal speed $v_{\phi}$ attained by the fastest rising apex of each $30-40$ kG Case TLfC flux tube at 0.95R as a function of emergence latitude.  The drag force from the prograde $\hat{v}_{\phi}$ acting on portions of the rising loop perpendicular to the mean azimuthal flow field has given the apex a positive azimuthal speed for nearly all 30-40 kG flux tubes that emerge at latitudes $\lesssim|25^{\circ}|$.  As the relevant components of the magnetic buoyancy and tension forces that drive parallel motion are comparatively smaller in 30-40 kG flux tubes, the trajectories of their rising apices are more easily accelerated equatorward by the application of differential rotation.  The TFT simulations of \citet{fan1994} similarly find that the application of a differential rotation profile reduced the emergence latitude of buoyantly rising loops in the Solar context.  


The 200 kG flux tubes initiated at 0.75R (both Case TLfC and TLsC) achieve low latitude emergence in a slightly different way than the 30-40 kG Case TLfC tubes.  The large excess of the outward forces acting on the flux tube compared to the inward forces quickly moves the tube outward to $\gtrsim0.9$R before horizontal equilibrium is achieved and the motion turns parallel to the rotation axis.  Subsequently, the emergence latitude deviates more significantly from what is predicted by Equation \ref{eqn:theta_parallel}.  This is visible in Figure \ref{fig:no_conv}b and Figure \ref{fig:choud_plot}c, where the motion parallel to the rotation axis of the 200 kG flux tube occurs much closer to the surface.  Modulation of the flux tube by the more vigorous, smaller scale convection in these layers perturbs the tube enough to allow portions to reach the upper boundary before the mean motion of the tube can move significantly poleward.  This effect is not strongly dependent on differential rotation, and likewise the initial $v_{\phi 0}$ established inside the flux tube.  This process also explains the reduced rise times of initially low latitude 200 kG flux tubes compared to the flux tubes rising in a quiescent convection zone in Figures \ref{fig:risetimes}b and \ref{fig:risetimes}d.    

\subsection{Relative Magnetic Pumping}
\label{sec:pump}


\begin{figure}

\vspace{.02\textwidth}
\includegraphics[scale=.42]{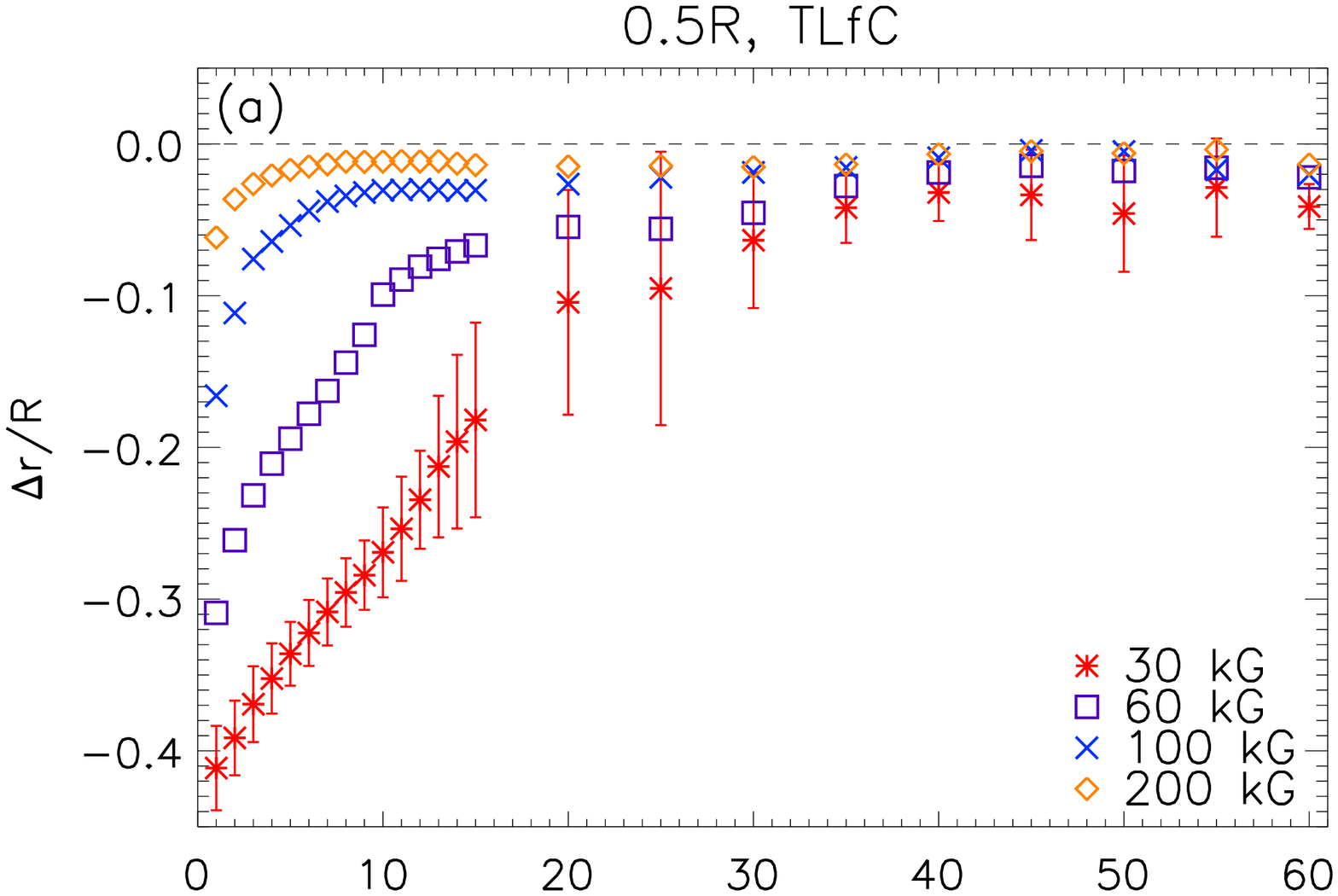} \\
\includegraphics[scale=.42]{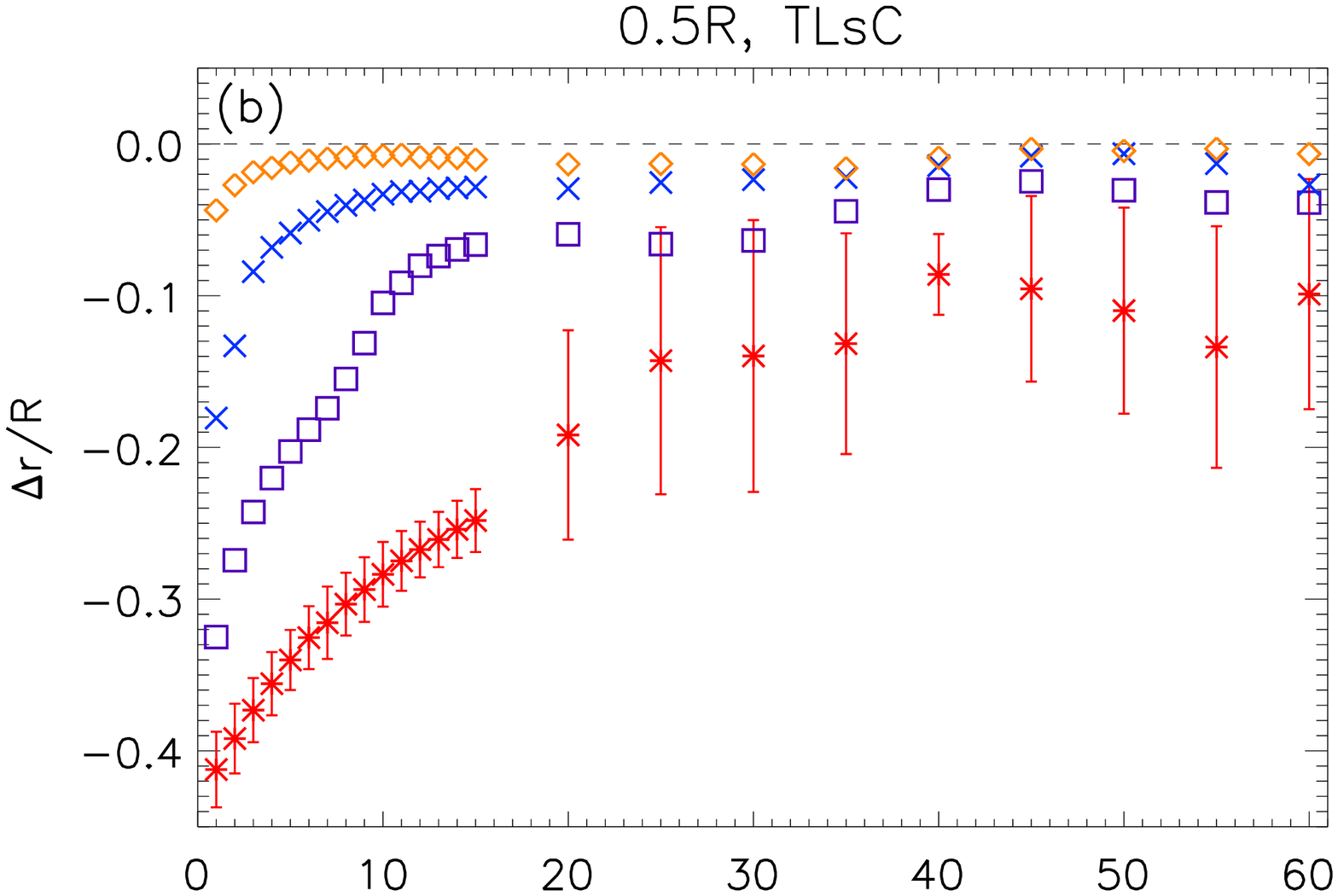} \\
\includegraphics[scale=.42]{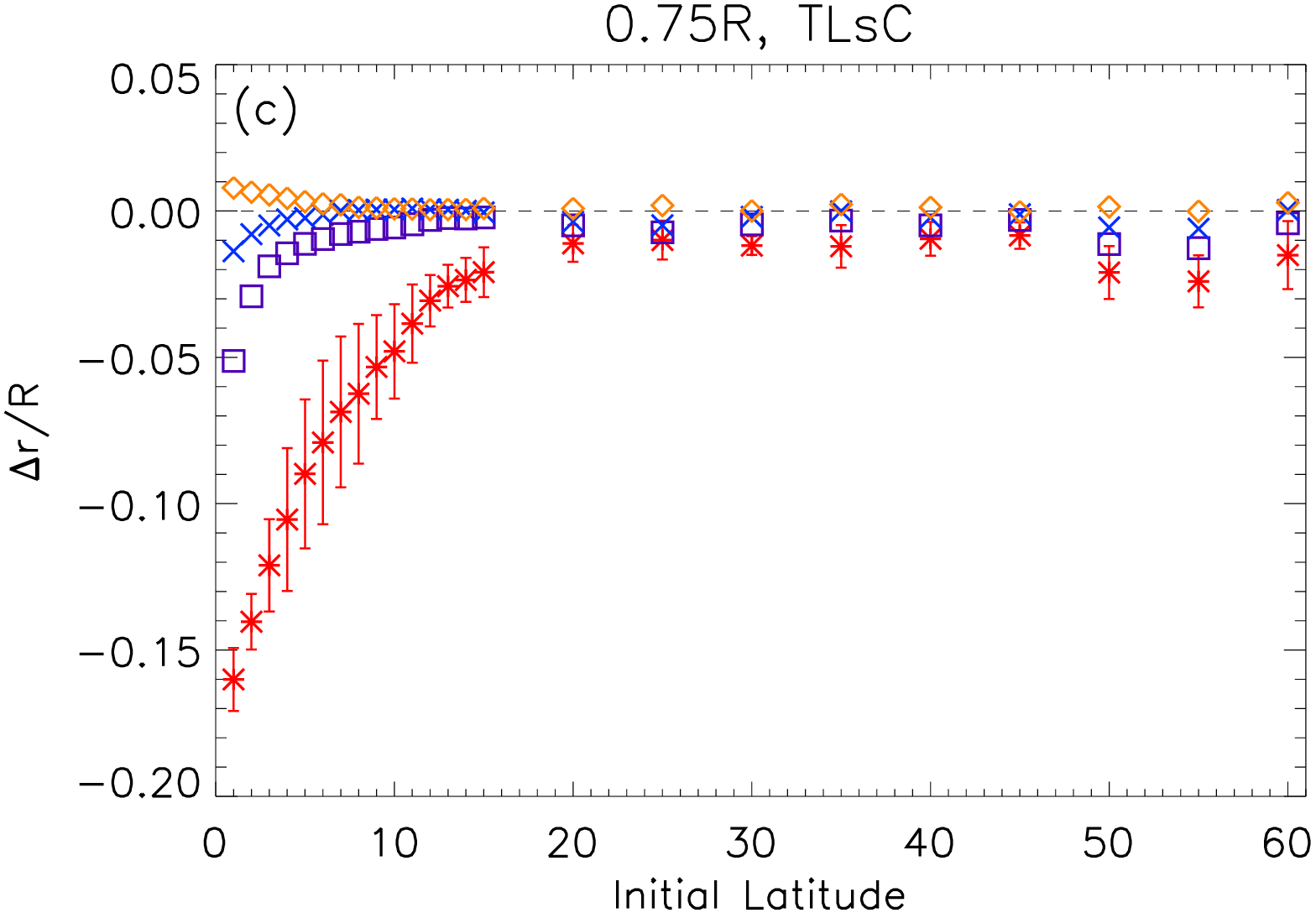}




\caption{Average relative pumping depth $\Delta r$ as a function of the absolute value of initial latitude for Case TLfC flux tubes initiated at 0.5R ({\bf a}) and TLsC flux tubes initiated at depths of 0.5R ({\bf b}) and 0.75R ({\bf c}).  Suppression of the global motion of the flux tube by convective downflows is more efficient in the deeper interior and at lower latitudes. Bars on the symbols represent the standard deviation, and are only shown for 30 kG flux tubes, which tend to show the largest spread in $\Delta r$ about the mean.}
\label{fig:deltar_avg}
\end{figure}

The flux tube properties reported in Section \ref{sec:diff} are reflective of only the fastest rising loop, assumed to be the progenitor of a representative starspot region.  However, a large portion of the originally axisymmetric ring may reside in deeper layers, as is evident by Figure \ref{fig:convection_tubes}.  Through interaction with a series of favorable flows, or by avoiding encounters with strong downflows, a buoyant loop may rise to the surface faster than a flux tube traveling through a quiescent interior.  Alternatively, the journey of even the fastest rising loop may be extended in time compared to the quiescent case due to pummeling of the flux tube by downflows.  

To quantify the ability of convective flows to suppress the mean motion of the flux tube, we calculate the average magnetic field weighted radial depth of the flux tube     
\begin{equation}
\langle r(t) \rangle =\frac{\int_{u_{0}}^{u_{N-1}} r(u,t) B(u,t) du }{\int_{u_{0}}^{u_{N-1}} B(u,t) du },
\end{equation} 
where $u_{j}=s/L=j/(N-1)$ for $j=0,..,N-1$ is the fractional arc length along the flux tube, $s$ is the length of the tube up to a mesh point $j$ from the origin mesh point, $L$ is the total length of the tube, and $N$ is the number of mesh points uniformly spaced along $L$.  This quantity $\langle r(t) \rangle$ captures the depth in the convection zone where the majority of the magnetic field of the flux tube resides.  It places less of a weight on the magnetic field in shallower depths which has decreased in strength as portions of the tube rise and expand.  Similar treatments for magnetic fields in 3D computational domains are employed in \citet[e.g.][]{tobias2001} and \citet{abbett2004}.

Figure \ref{fig:pumping} compares as a function of time the maximum radial position $r_{c}$ of representative Case TLfC and TLsC flux tubes (solid lines) and the corresponding magnetic field weighted radial position $\langle r \rangle$ (dash-triple-dotted lines).  In addition, we show the cross-sectional radial position $r_{nc}$ of the axisymmetric flux tube (dashed lines) rising through the quiescent interior with the same initial conditions.  This figure indicates that the majority of the flux tube is confined to deeper layers than the fastest rising portion of the tube, while $\langle r \rangle$ may deviate more or less from $r_{nc}$ according to the magnetic field strength $B_{0}$, depth $r_{0}$, latitude $\theta_{0}$, and angular velocity contrast.

The deviation of $\langle r \rangle$ from the depth $r_{nc}$ of the axisymmetric flux tube rising in a quiescent convection zone is an expression of the effectiveness of magnetic pumping.  To quantify this, we calculate a \emph{relative pumping depth} for each individual flux tube simulation $\Delta r = \langle r(t_{min}) \rangle -r_{nc}(t_{min})$, where $t_{min}=\min[t(\max(r_{nc})), t(\max(r_{c}))]$.  In other words, we record the difference between $\langle r \rangle$ and $r_{nc}$ at the elapsed time corresponding to whichever flux tube reaches the simulation upper boundary first, either the flux tube evolving in convective flows (i.e. $r_{c}$) or the flux tube evolving without convective motions (i.e. $r_{nc}$).  For example, the 30 kG Case TLf flux tube depicted in Figure \ref{fig:pumping}a reaches the simulation upper boundary in $\sim$445 days.  This is $\sim$45 days faster than the same flux tube subject to convective flows (Case TLfC).  In this instance, $t_{min}\sim445$ days, represented by the vertical dashed line.  At $t_{min}$, $\Delta r \sim$ 0.620R - 0.950R $= -0.330$R, indicating that the Case TLfC flux tube has been pumped downward by convection: i.e., $\langle r \rangle$ is increasing substantially slower than $r_{nc}$.  In Figure \ref{fig:pumping}b, the 30 kG Case TLsC tube reaches the surface in $\sim$45 days.  This is $\sim$10 days faster than the equivalent flux tube evolving without convective flows (Case TLs).  In this case, $t_{min}\sim45$ days, with $\Delta r \sim$ 0.858R - 0.873R $= -0.015$R.  The average $\Delta r$ for our simulations is shown in Figure \ref{fig:deltar_avg}.  As in Figures \ref{fig:risetimes} and \ref{fig:theta_em}, each symbol in the plot represents the quantity for all flux tubes initiated at $|\theta_{0}|$, averaged over the three ensembles we perform.  We emphasize that in the calculation of $\Delta r$, we are always comparing flux tubes with the same $v_{\phi 0}$.  

Taking Figures \ref{fig:pumping} and \ref{fig:deltar_avg} together, it is clear in the deep interior that the relative magnetic pumping is more efficient for flux tubes of weaker magnetic field strengths and lower initial latitudes.  Figure \ref{fig:deltar_avg} demonstrates this comprehensively, especially when comparing the Case TLsC flux tubes initiated at both 0.5R and 0.75R in panels \ref{fig:deltar_avg}b and \ref{fig:deltar_avg}c, respectively.  At depths of 0.5R, the critical magnetic field strength at which the magnetic buoyancy roughly equals the downward drag force from radial convective motions is $B_{c}\sim30$ kG.  At weaker $B_{0}$, there is a continuous tug-of-war between buoyancy and convective motions until a loop is lucky enough to rise to the surface without being pummeled back downward by convection.  

For the plots of $r_{c}$ in Figure \ref{fig:pumping}, we emphasize that we are always tracing the portion of the flux tube that has the largest radial distance from the star's center.  As the tube evolves, various loops will develop over the course of the simulation that subsequently are pushed downward by convective flows.  As a result, we are not tracking a single loop from the beginning of the simulation to termination at the upper boundary.  Even at magnetic field strengths close to $B_{c}$, the average magnetic field weighted depth $\langle r \rangle$ increases with time owing to the lack of a stably stratified region to help anchor the tube, and the unbalanced vertical and poleward force components acting on the flux tube as a whole.

The reduced pumping at higher latitudes is in part a consequence of the larger poleward acceleration due to the smaller radius of curvature there.  Convective flows are not strong enough to retard the motion as efficiently.  The relative pumping depth is larger at higher latitudes for $30-40$ kG flux tubes initiated at 0.5R evolving in the slow differential rotation profile (Case TLsC, Fig. \ref{fig:deltar_avg}c) as compared to those evolving in the fast differential rotation profile (Case TLfC, Fig. \ref{fig:deltar_avg}a).  This indicates that the nature of the differential rotation profile acting on the flux tube as a whole also plays some role in magnetic pumping.  


At initial depths of 0.75R, any substantial pumping is only seen for the lower latitude, lower magnetic field strength Case TLsC flux tubes shown in Figure \ref{fig:deltar_avg}c.  For the Case TLfC flux tubes initiated at this height, by contrast, the initial imbalance of the horizontal forces brings the tubes horizontally outward into shallower layers very quickly (see discussion in Sec. \ref{sec:dynamic}).  Magnetic pumping here is not strong enough to suppress this initial outward motion.  As a result, the mean motion of the these flux tubes evolve similarly to those that rise through a quiescent convection zone, with average $|\Delta r/R|$ values never greater than 0.04 (and subsequently not shown here).  That is not to say that all portions of the flux tube evolve in a similar fashion.  As discussed in Section \ref{sec:diff}, some buoyantly rising loops may escape to the surface much quicker and more radially than their counterparts evolving without convection.  A general trend emerges: magnetic pumping is more efficient for weaker magnetic field strengths, lower latitudes, in the deeper interior, and smaller angular velocity contrasts.


\section{Conclusions and Perspectives}
\label{sec:conclude}

We have presented the results from simulations of thin flux tubes embedded in a rotating spherical domain of fluid motions representative of a 0.3M$_{\odot}$ fully convective star.  Our simulations are meant to represent how coherent bundles of initially toroidal magnetic fields with core strengths of $B_{0} \gtrsim B_{c}$, where $B_{c} \sim 20-30$ kG, and moderate magnetic flux of $\sim$$10^{21}-10^{22}$ Mx may behave as they traverse the convection zone, interacting with local and mean flows.  We recapitulate our findings in what follows, comment on the comparison to previous simulations and observations of active regions on M-dwarfs, and discuss the assumptions made in our model.  

Flux tubes initially in thermal equilibrium (TEQ) rise as axisymmetric rings in a quiescent interior.  The early motion of the flux tube is largely dominated by the initial horizontal force imbalance, with the evolution continually adjusting until the sum of the inward and outward directed components of the magnetic buoyancy, tension, Coriolis force, and drag force reaches equilibrium.  Once a horizontal equilibrium is reached, the motion of the ring turns parallel to the rotation axis, driven by the unbalanced vertical component of buoyancy and poleward components of magnetic tension and the Coriolis force.  A similar behavior is found for TFT simulations of axisymmetric flux rings rising through a quiescent Solar convection zone \citep[e.g.][]{choud1987}.  The effect of radiative heating on the flux tubes, though included in our simulations, is minimal.


To our knowledge, TFT simulations in a fully convective star have only been considered here and in \citet{browning2016}.  Traditional TFT models tend to assume that the dynamo mechanism generates toroidal flux tubes at the interface between the radiative interior and the convection zone, often assumed to be in mechanical force equilibrium (MEQ) and neutral buoyancy \citep[e.g.][]{caligari1995,weber2011,granzer2000,holzwarth2001}.  However, recent simulations of global-scale dynamo action in spherical shells \citep[e.g.][]{nelson2014} have demonstrated that, at least in some parameter regimes, buoyant magnetic loops can be built by a distributed dynamo without a tachocline region.  Rather than achieving neutral buoyancy and MEQ, as is argued for flux tubes built by an interface dynamo, it is more likely that flux tubes built by a distributed dynamo achieve a state closer to that of TEQ.  To facilitate some level of comparison between these initial condition assumptions, we also perform some simulations where flux tubes initially in MEQ rise through a quiescent convection zone. Similar to TFT models in stars with small radiative cores \citep[e.g.][]{granzer2000,holzwarth2001}, we find that most of these flux tubes slip poleward, driven by a magnetic tension force that is dominant compared to buoyancy.  However, strongly super-equipartition flux tubes can develop low order ($m=1-2$) unstable modes if the magnetic field strength and radius of curvature is large enough.

The Solar archetype of buoyantly rising, loop-shaped magnetic structures is not achievable in the interior of fully convective stars if the initially toroidal flux tube is assumed to be in TEQ, as we show in Section \ref{sec:dynamic}.  However, the addition of a time-varying convective velocity field modulates the axisymmetric flux ring, promoting rising loops.  All flux tubes we study initiated in the deep interior have apices (i.e. the portion reaching the simulation upper boundary first) that rise almost exactly parallel to the rotation axis, following the relationship given in Equation \ref{eqn:theta_parallel}.  However, when subjected to a strong differential rotation profile with an angular velocity contrast comparable to the Sun, and likewise built co-rotating with the local azimuthal flow, flux tubes of a few times $B_{c}$ ($30-40$ kG) initiated at low latitudes ($\lesssim10^{\circ}-15^{\circ}$) and in shallower depths of 0.75R are able to emerge at latitudes significantly lower than what Equation \ref{eqn:theta_parallel} predicts, and even at the equator.  The strong prograde $v_{\phi 0}$ inside the tube initially causes it to move into shallower layers via induced Coriolis forces.  The differential rotation in these layers supplies angular momentum to the legs of the rising loop, facilitating lower latitude emergence (as seen in \citet{fan1994} in the Solar context).  A reduced rise time arises, in part, from this action as well, but is also a result of the more radial trajectory, requiring the flux tube apex to traverse a shorter distance to the surface.  When subjected to a slower differential rotation profile closer to that of a rigid rotator, and more commensurate with observations of M-dwarfs, only strongly super-equipartition flux tubes of 200 kG are able to emerge in the near-equatorial region.  The large magnetic buoyancy brings the tube horizontally outward to very near the surface, where the stronger downflows of smaller spatial scale modulate the flux tube to promote buoyant loops before the mean motion of the flux tube can turn poleward.

With our simulations, we can also assess the efficiency of magnetic pumping.  By avoiding strong downflows, or encountering a series of favorable mean and local flows, a buoyant loop may be boosted to the surface faster than if the tube travelled through a quiescent interior.  Alternatively, the mean motion of the flux tube may be suppressed by continual pummeling of the flux tube by downflows.  As expected, flux tubes of weaker magnetic field strength are pumped to a greater degree \citep[e.g.][]{tobias2001,abbett2004}.  For flux tubes of the same magnetic field strength, we find that magnetic pumping is most efficient in the deeper interior and nearer the equator.  This is partly a consequence of reduced flux tube buoyancy in the deeper interior and a reduced magnetic tension at lower latitudes (compared to higher latitudes at the same radial depth).  


Our results suggest that emerging flux tubes, if produced in the interior, could plausibly account for the appearance of starspots at mid-to-high latitudes. But what of lower latitude spots, especially those appearing near the equator?  There are a few possible scenarios to explain the appearance of very low latitude starspots: (1) the magnetic field strength $B_{0}$ of progenitor flux tubes are strongly super-equipartition, (2) the differential rotation profile in the interior is much stronger than what is observed at the surface, (3) the tubes eventually emerging at the surface are generated in much shallower regions than what is accessible through our TFT simulations given our choice of $B_{0}$ and magnetic flux of $\sim10^{21}-10^{22}$ Mx, (4) the active regions are a grouping of much smaller magnetic flux bundles, with $\Phi < 10^{21}$ Mx.  To truly elucidate the processes responsible for equatorial and low-latitude starspots in fully convective star, more work is needed.  

We speculate that the longitudinal extent of active regions that may be produced by our simulations is limited by the size of the cellular convective structures in the upper convection zone, where radial downflows have the strongest amplitudes.  Similar to the TFT simulations of \citet{weberapj2013} in the Solar convection zone, strong downflow lanes at the edges of giant cells might contribute to a preferential longitudinal emergence of active regions.  Addressing the `spottedness' of the star, in particular the percentage of starspot coverage on the surface at any time, is beyond the scope of the simulations presented here.

Like all simulations of stellar convection and magnetic flux emergence, we have made a number of simplifications in our modeling.  Many of these stem from the fundamental assumptions and numerical requirements of the anelastic and TFT approximations.   Arguably, the most significant assumption we have made is that dynamo action in a fully convective star builds coherent, individual flux tubes in a toroidal geometry distributed throughout the bulk of the interior.  The true nature of dynamo generated magnetic fields in the interior of any star is still largely unknown.  In fully convective stars, it may well be the case that the poloidal field plays a role in flux emergence as well, rather than the toroidal field alone.  


The TFT approximation also does not resolve the cross-section of the flux tube, assuming that it always remains circular with a radius $a=(\Phi/B\pi)^{1/2}$.  As the tube rises, it expands and attains greater speeds.  A flux tube moving transversely through a fluid will feel a pressure excess on the leading and trailing surfaces, with a pressure deficit on the sides.  Such a pressure difference could flatten the flux tube into a ribbon-like shape, reducing the rise speed \citep[e.g.][]{parker1975}.  The tube may fragment or become shredded by convective motions, perhaps developing a more umbrella-shaped cross-section \citep[e.g.][]{schuessler1979,fan1998}.  This could effectively be captured in the drag force of the TFT model by varying the drag coefficient as the tube rises.  We plan to investigate this effect in the future in the Solar context.

We have employed an angular velocity $\Omega_{0}=$2.6\e{-6} rad s$^{-1}$, comparable to the Sun.  For a star of radius 2.0\e{10} cm, this implies a rotational velocity $v_{rot}=0.5$ km s$^{-1}$, below the current $v \sin{i} \sim 2$ km s$^{-1}$ detection limit for Doppler broadening \citep[e.g.][]{delfosse1998,browning2010,reinersaj2012}.  From an observational standpoint, our simulations are essentially non-rotating, with some M-dwarf rotational velocities exceeding $v \sin{i} \sim 20$ km s$^{-1}$ \citep[e.g.][]{jenkins2009}.  Investigation of dynamo action and flux emergence in more rapidly rotating, fully convective stars have been planned for the future.  Broadly, we anticipate that more rapidly rotating objects would show an even stronger tendency towards flux emergence near the poles.  Rapid rotation may also alter the `giant cell' convective structure pattern \citep[e.g.][]{brown2008}. 

Our choice of flux tube initial conditions and the treatment of the external velocity field could, in some cases, contribute to our results in a non-negligible way.  Foremost among these is assuming a density deficit inside the tube prescribed by the condition of TEQ, resulting in a substantial buoyancy force compared to the neutrally buoyant state of flux tubes in MEQ.  It may well be the case that flux tubes are built in a state somewhere between these extremes.  The assumption that flux tubes are built co-rotating with the local plasma, initializing a flow $v_{\phi 0}$ inside the tube, also plays a role in the subsequent force evolution.  Modeling flux tubes with a smaller cross-sectional radius $a$ than the chosen $\sim$2\e{8} cm could result in more examples of low latitude emergence, as such tubes are advected more strongly by convective flows.  It is also likely that the strong magnetic fields we consider could quench the differential rotation, reducing the angular velocity contrast, similar to that of Case Cm in \citet{browning2008}.  The method we use to treat the external velocity field, discussed in Section \ref{sec:ASH}, does not take into account corresponding changes made to the time-varying radial and longitudinal velocity field due to dynamo action.  However, we point out that the amplitudes of the radial velocity field and nature of the giant-cell convection in the hydrodynamic simulation used here is commensurate with Case Cm.  The amplitude of the meridional circulation established in either case likely contributes little to flux tube evolution.   
   
The modeling approach we take is advantageous in that we can prescribe a number of parameters to see how flux tubes initiated in different regions of the star with various magnetic field strengths may behave.  While our tubes are not generated self-consistently, we can evaluate many possible flux tubes that may exist, and examine how they might behave under certain external conditions.  Although observations of M-dwarfs are increasing in number, detailed and long term observations of active regions on these stars is limited.  Our understanding of stellar magnetism is largely driven by what we have observed on the Sun.  Ideally, we would hope to retrieve information about whether uniform starspot coverage is the norm, if active regions are mostly bipolar as they are on the Sun, and if so, whether they exhibit tilting toward the equator following Joy's Law, and/or opposite polarities in opposite hemispheres following Hale's law.  Such detailed observations of individual active regions on any star other than the Sun are likely years in the future.  Until then, we turn to theory and simulations to guide our knowledge of the in-depth operation of stellar dynamos.  The TFT simulations we have presented here serve to complement existing 3D dynamo simulations of fully convective stars, providing a link between dynamo-generated magnetic fields, fluid motions, and starspots.

\acknowledgments
This work was supported by the European Research Council under ERC grant agreements No. 337705 (CHASM) and by a Consolidated Grant from the UK STFC (ST/J001627/1). Some of the calculations for this paper were performed on the DiRAC Complexity machine, jointly funded by STFC and the Large Facilities Capital Fund of BIS, and the University of Exeter supercomputer, a DiRAC Facility jointly funded by STFC, the Large Facilities Capital Fund of BIS and the University of Exeter.  We thank Isabelle Baraffe for providing the 1D stellar structure model used in this paper, and for offering comments on an earlier version of this manuscript.  We also thank the referee for a thorough reading of our paper and providing many helpful comments, improving the clarity of our paper.

\clearpage

\end{document}